\documentclass[11pt]{article}
\usepackage{axodraw}
\usepackage{latexsym}
\makeatletter \@addtoreset{equation}{section} \makeatother

\newtheorem{theorem}{Theorem}[section]
\newtheorem{corollary}{Corollary}

\newtheorem{lemma}[theorem]{Lemma}
\newtheorem{proposition}[theorem]{Proposition}
\newtheorem{definition}{Definition}

\newcommand{\tre}{\check{\EE}_{(d,H)}}
\newcommand{\treh}{\check{\EE}_H}
\newcommand{\NT}{{N}}
\newcommand{\dnth}{d^{\NT}\theta\;}
\newcommand{\te}{\theta}
\newcommand{\cas}{{\mbox{\footnotesize$\cal S$}}}
\newcommand{\dth}{\pa_\theta}
\newcommand{\tQ }{{\tilde Q}}

\newcommand{\hO}{{\hat\Omega}}
\newcommand{\hA}{{\hat A}}

\newcommand{\hd}{{\hat d}}
\newcommand{\hF}{{\hat F}}

\newcommand{\trO}{\check{\Omega}}
\newcommand{\trd}{\check{d} \,}
\newcommand{\trde}{\check{\delta} \,}

\newcommand{\trfi}{\check{\phi}}
\newcommand{\trpsi}{\check{\psi}}
\newcommand{\exfi}{\tilde{\phi}}
\newcommand{\trfipr}{\check{\phi}{}\,'\,}
\newcommand{\trpsipr}{\check{\psi}{}\,'\,}
\newcommand{\exfipr}{\tilde{\phi}{}'\,}
\newcommand{\dpr}{{\check\delta}{}^{\prime}}
\newcommand{\exdpr}{{\tilde\delta}{}^{\prime}}

\newcommand{\truncation}{{\,^{({\rm tr})}}}
\newcommand{\trunpr}{{\,^{({\rm tr'})}}}

\def\ii{\'\i}

\def\coes{\c c\~oes}

\def\ii{\'\i}

\def\coes{\c c\~oes}

\def\ftoday{{\sl {Le \number\day \space\ifcase\month
\or janvier\or f\'evrier\or mars\or avril\or mai \or juin\or juillet\or
ao\^ut\or septembre\or octobre \or novembre \or d\'ecembre\fi\space
\number\year}}}
\def\ptoday{{\sl {\number\day \space de\space \ifcase\month
\or janeiro\or fevereiro\or mar{\c c}o\or abril\or maio \or junho\or
julho\or agosto\or setembro\or outubro \or novembro \or dezembro\fi\space
de\space \number\year}}}
\def\gtoday{{\sl {Den \number\day. \ifcase\month
\or Januar\or Februar\or M\"arz\or April\or Mai \or Juni\or Juli\or
August\or September\or Oktober \or November \or Dezember\fi\space
\number\year}}}
\def\today{{\sl {\ifcase\month
\or January\or February\or March\or April\or May \or June\or July\or
August\or September\or October \or November \or December\fi
\space\number\day,\space
                                            \number\year}}}
\newcommand{\journal}[4]{{\em #1~}#2\,(#3)\,#4}

\newcommand{\ijmp}{\journal {Int. J. Mod. Phys.}}

\newcommand{\pr}{\journal {Phys. Rev.}}

\newcommand{\cmp}{\journal {Commun. Math. Phys.}}

\newcommand{\np}{\journal {Nucl. Phys.}}
\newcommand{\npproc}{\journal {Nucl. Phys. B (Proc.Suppl.)}}
\newcommand{\pl}{\journal {Phys. Lett.}}

\newcommand{\prep}{\journal {Phys. Rep.}}

\newcommand{\jdiffgeom}{\journal {J. Diff. Geom.}}
\newcommand{\topology}{\journal {Topology}}
\newcommand{\eurphys}{\journal {Eur. Phys. J.}}

\renewcommand{\a}{\alpha}

\renewcommand{\d}{\delta}         \newcommand{\D}{\Delta}
\newcommand{\e}{\epsilon}

\newcommand{\la}{\lambda}        
\newcommand{\m}{\mu}

\newcommand{\om}{\omega}         \newcommand{\OM}{\Omega}

\newcommand{\f}{{\phi}}           \newcommand{\F}{{\Phi}}
\newcommand{\vf}{{\varphi}}
\newcommand{\XI}{\XI}

\newcommand{\EE}{{\cal E}}

\newcommand{\ZZ}{{\cal Z}}
\newcommand{\es}{\\[3mm]}

\newcommand{\superint}{\dint\kern -.85em \raise -.50ex\hbox{\Large S}\;}
\newcommand{\sla}{\raise.15ex\hbox{$/$}\kern -.57em}
\newcommand{\Sla}{\raise.15ex\hbox{$/$}\kern -.70em}

\def\Lp{\displaystyle{\biggl(}}
\def\Rp{\displaystyle{\biggr)}}

\newcommand{\lp}{\left(}\newcommand{\rp}{\right)}
\newcommand{\lc}{\left[}\newcommand{\rc}{\right]}
\newcommand{\lac}{\left\{}\newcommand{\rac}{\right\}}

\newcommand{\complex}{{\kern .1em {\raise .47ex
\hbox {$\scriptscriptstyle |$}}
    \kern -.4em {\rm C}}}
\newcommand{\real}{{{\rm I} \kern -.19em {\rm R}}}
\newcommand{\rational}{{\kern .1em {\raise .47ex
\hbox{$\scripscriptstyle |$}}
    \kern -.35em {\rm Q}}}
\renewcommand{\natural}{{\vrule height 1.6ex width
.05em depth 0ex \kern -.35em {\rm N}}}

\newcommand{\tr}{{\rm {Tr} \,}}
\newcommand{\half}{\dfrac{1}{2}}
\newcommand{\pa}{\partial}

\newcommand{\dpad}[2]{{\displaystyle{\frac{\partial #1}{\partial
#2}}}}

\newcommand{\dfrac}[2]{{\displaystyle{\frac{#1}{#2}}}}
\newcommand{\dsum}[2]{\displaystyle{\sum_{#1}^{#2}}}
\newcommand{\dint}{\displaystyle{\int}}

\newcommand{\ie}{{{\em i.e.},\ }}

\newcommand{\twiddle}{\lower.9ex\rlap{$\kern -.1em\scriptstyle\sim$}}
\newcommand{\qed}{{\hfill$\Box$}}

\newcommand{\equ}[1]{(\ref{#1})}
\newcommand{\eq}{\begin{equation}}
\newcommand{\eqn}[1]{\label{#1}\end{equation}}
\newcommand{\eea}{\end{eqnarray}}
\newcommand{\eqa}{\begin{eqnarray}}
\newcommand{\eqan}[1]{\label{#1}\end{eqnarray}}
\newcommand{\ba}{\begin{array}}
\newcommand{\ea}{\end{array}}
\newcommand{\eqac}{\begin{equation}\begin{array}{rcl}}
\newcommand{\eqacn}[1]{\end{array}\label{#1}\end{equation}}

\begin{document}

{\hfill\parbox{45mm}{{
UFES-DF-OP2005/1
}} \vspace{3mm}

\begin{center}
{\LARGE\bf    Observables in Topological Yang-Mills Theories\es
With Extended Shift Supersymmetry}
\end{center}
\vspace{3mm}

\begin{center}{\large
Clisthenis P. Constantinidis$^{a}$,
Olivier Piguet$^{a}\footnote{Supported
in part by the Conselho Nacional
de Desenvolvimento Cient\'{\i}fico e
Tecnol\'{o}gico CNPq -- Brazil.}$ and
Wesley Spalenza$^{b,c,d,1}$ } \vspace{1mm}

\noindent 
$^{a}$Universidade Federal do Esp\'{i}rito Santo (UFES) Brazil\\
$^{b}$Scuola Internazionale Superiore di Studi Avanzati (SISSA) - Trieste, Italia\\
$^{c}$Centro Brasileiro de Pesquisas F\ii sicas (CBPF) - RJ, Brazil\\
$^{d}$Grupo de F\'{i}sica Te\'{o}rica Jos\'{e} Leite Lopes (GFT)
\end{center}
\vspace{1mm}

{\tt E-mails: clisthen@cce.ufes.br, opiguet@yahoo.com, spalenza@sissa.it}

\begin{abstract}
We present a complete classification, at the classical level, 
of the observables of topological Yang-Mills theories 
with an extended shift supersymmetry  of $\NT$ generators, 
in any space-time
dimension. The observables are defined as the 
Yang-Mills BRST cohomology classes of
shift supersymmetry invariants. 
 These cohomology classes turn out to be solutions of an $\NT$-extension 
of Witten's equivariant cohomology.
This work generalizes results known in
the case of shift supersymmetry with a single generator.
\end{abstract}

\section{Introduction}

The prototype for topological theories of Witten's type
is the four dimensional topological Yang-Mills theory of 
Witten~\cite{witten-donald,baulieu-singer,bbrt-pr}, whose quantum observables are the
Donaldson invariants~\cite{witten-donald,donaldson}. This model is
characterized by a shift invariance, or ``shift supersymmetry'',
generated by a single scalar fermionic charge, which is 
interpreted as the BRST invariance describing the nonphysical character  
of the connection, with the result that only global ``observables'', namely
the Donaldson invariants, are present. Generalizations to
supersymmetry (SUSY) with  $\NT=2$ or more generators 
where already proposed some years
ago in~\cite{yamron,blau-thom-96,marculescu} and more recently 
in~\cite{braga,geyer,nos,tese-Wesley}. 
The construction of Lagrangian models for the gauge fixing 
of the shift supersymmetries may be found in~\cite{blau-thom-96} and, 
for arbitrary $\NT$ and arbitrary
space-time dimension, in~\cite{nos,tese-Wesley}.
Most of these constructions are based on a superspace formalism for 
shift-SUSY introduced first in~\cite{horne}. 
There, gauge invariance is formulated in terms 
of superfields. The Faddeev-Popov ghosts being superfields, 
supergauge invariance represents invariances with respect to a 
supermultiplet of local symmetries. However all these local
invariances,
except the original gauge invariance,
may be fixed algebraically in a manner very similar to the Wess-Zumino
(WZ) gauge fixing of usual supersymmetric gauge theories~\cite{1001}. 
The equivalence of
the superspace theory and of the original one, 
e.g. that of Witten in the $\NT=1$ case, is made explicit in this
WZ like gauge~\cite{blau-thom-96,boldo,nos}.

The problem of the characterization of the observables of such theories
was defined by Witten~\cite{witten-donald,ouvry} as the computation of 
``equivariant cohomology'', i.e. the cohomology of the shift-SUSY
generator in the space of the gauge invariant local functionals of the fields.
It was shown in~\cite{boldo}  that this problem is almost equivalent
to the presumably more tractable one of calculating, in the superspace formalism, 
the cohomology of the BRST operator associated
with superspace gauge invariance, in the space of shift-SUSY invariant local
functionals. In fact, this equivalence 
is exact up to  solutions which are obviously trivial in the sense of Witten's
equivariant cohomology. 

Our purpose is to characterize, for general $\NT$ and general 
space-time dimension, 
and in a formal classical set-up, 
all the observables defined as solutions of the BRST cohomology for SUSY
invariant objects. We will also show that these solutions -- up to
some of them which turn out to be obviously trivial -- are equivalent 
to solutions
of an equivariant cohomology, defined in the WZ-gauge as a
generalization of the $\NT=1$ definition of Witten. 

Section~\ref{ext-susy} presents an introduction to the superspace formalism, 
with superfields, superforms, supergauge invariance, superconnection, 
superghosts and BRST symmetry.
The WZ-gauge fixing is recalled in 
Section~\ref{observables-equiv},
where also a generalized definition of equivariant cohomology is proposed.
The problem of finding the observables and its solution are explained 
in Section~\ref{observables-brst}. In Section \ref{witten-observ},
we write down the general result in the WZ-gauge, show that the
integrants of the observables obey a set of generalized Witten's descent
equations and that they are nontrivial in the sense of equivariant
cohomology.
Our conclusions are presented in Section~\ref{conclusion}. Superspace 
conventions and notations are given in Appendix~\ref{app-a}.
Some useful proposition on the relative cohomologies of a general set of
$n$ coboundary operators, needed in the main text,
 are stated and proved in Appendix~\ref{app-b}.
Appendix~\ref{app-c} contains the proof of another proposition of
a more technical character.
 The WZ-gauge is recalled in Appendix~\ref{app-d}, where a
one-to-one correspondence between the fields in this gauge and 
a set of covariant superfields is constructed.

This paper may be viewed as a continuation of both papers~\cite{boldo} 
and~\cite{nos}, the first reference dealing with the problem of the 
observables in the case $\NT=1$, and the second one presenting an
introduction to the superspace formalism for $\NT>1$ and the reduction to
the Wess-Zumiono gauge.
Prelimimary results of the present work were presented 
in~\cite{tese-Wesley}.

\section{$\NT$-Extended Supersymmetry}\label{ext-susy}

``Shift supersymmetry'' may describe the
gauge fixing of gauge field configurations with null curvature,
or alternatively with selfdual curvature. It appeared originally in
Donaldson-Witten model~\cite{witten-donald,donaldson},
with one supersymmetry generator in four dimensional space-time.
Generalizations of it for more than one supersymmetry
generators and for any space-time dimension were
described in \cite{horne,blau-thom-96,boldo, nos}, where a superspace
formalism has been developed.  The purpose of this section is to review
the formalism and fix the notation.

\subsection{$\NT$ Superspace formalism}

$\NT$ supersymmetry is generated by the fermionic charges
$Q_I$, $I=1,\dots,\NT$ obeying the Abelian
superalgebra  \footnote{The bracket is here an anti-commutator. Troughout
this paper brackets will denote either commutators or anticommutators,
according to the statistics of their arguments.}
\eq
[Q_I,Q_J] = 0\ ,
\eqn{s-agebra}
commuting with the space-time
symmetry generators and the gauge group generators. The gauge group is some
compact Lie group.

A representation of supersymmetry is provided by superspace, a
supermanifold with $D$  bosonic and $\NT$
fermionic dimensions\footnote{Notations and conventions on
superspace are given in Appendix \ref{app-a}.}.
The respective coordinates are denoted by
$(x^{\mu }$, $\mu=0,\dots,D-1)$, and  ($\theta ^{I}$,  $I=1,\dots,\NT)$.
A superfield is by definition a superspace function  $F(x,\theta )$
which transforms as
\begin{equation}
Q_{I}F(x,\theta ) = \partial_{I}F(x,\theta )
\equiv  \frac{\partial }{\partial \theta ^{I}}F(x,\theta )
\eqn{s-field-transf}
under an infinitesimal supersymmetry transformation.

An expansion in the coordinates $\theta ^{I}$ of a generic superfield
reads
\begin{equation}
F(x,\theta )=f(x)+\dsum{n=1}{{\NT}}\frac{1}{n!}\theta
^{I_{1}}\cdots \theta ^{I_{n}}f_{I_{1}\cdots I_{n}}(x)
\eqn{s-field-exp}
where the space-time fields $f_{I_{1}\cdots I_{n}}(x)$ are completely
antisymmetric in the indices $I_{1}\cdots I_{n}$. We recall that all fields
(and superfields) are Lie algebra valued.
 These fields and superfields may be generalized to $p$-forms and
superfield $p$-forms
\eq
\OM_p(x,\theta )=\om_p(x)+\dsum{n=1}{{\NT}}\frac{1}{n!}\theta
^{I_{1}}\cdots \theta ^{I_{n}}\om_{p,\,I_{1}\cdots I_{n}}(x)\ .
\eqn{s-field-form-exp}
In \equ{s-field-exp} or \equ{s-field-form-exp}, the components $n\ge1$ are
SUSY transforms of the lowest component. This may be viewed explicitly
through the identity
\begin{equation}
\Omega_p(x,\theta)=\,\,\exp \{\theta
^{I}Q_{I}\}\,\omega _p(x)\,\,=\,\,\,\sum_{n=0}^{\NT}\frac{1}{n!}
\theta ^{I_{1}}\cdots \theta ^{I_{n}}\,\,Q_{I_{n}}\cdots 
Q_{I_{1}}\,\,\omega_p(x)   \ ,
\label{sfieldform}\end{equation}
which holds due to the easily checked superfield property
$\pa_I\OM_p = Q_I\OM_p$ and the fact that a superfield is uniquely
determined by its $\theta=0$ component.

We shall also deal with superforms. 
  A $q$-superform may be written as
\eq
\hO_q = \sum_{k=0}^{q}  
\OM_{q-k;\,I_1\cdots I_k} d\theta^{I_1}\cdots d\theta^{I_k}\ ,
\eqn{s-form}
where the coefficients $\OM_{q-k;\,I_1\cdots I_k}$ are (Lie algebra valued)
superfields which are space-time forms of degree $(q-k)$.  
They are
completely symmetric in their indices since, the
coordinates $\theta $ being anti-commutative, the
differentials $d\theta^I $ are commutative.
The superspace exterior derivative is defined as
\begin{equation}
\hd=d+d\theta ^{I}\partial _{I}\ ,\quad d=dx^{\mu }\partial _{\mu }\ ,
\end{equation}
and is nilpotent: $\hd\;^{2}=\;0$.

The basic superfield of the theory is the superconnection $\hA$,
a $1$-superform:
\begin{equation}
\hA=A+E_{I}d\theta ^{I}  \ ,
\label{superconexao}\end{equation}
with $A=A_{\mu }(x,\theta )dx^{\mu }$ a $1$-form superfield and
$E_{I}=E_{I}(x,\theta )$ a $0$-form superfield. The
superghost $C(x,\theta)$ is a $0$-superform.
 We expand the components of the superconnection (\ref{superconexao}) as
\begin{equation}
A=a(x)+ \dsum{n=1}{{\NT}}\frac{1}{n!}\theta
^{I_{1}}\cdots \theta ^{I_{n}}a_{I_{1}\cdots I_{n}}(x)\ ,
\label{expansion-A}\end{equation}
where the 1-form $a$ is the gauge connection, and the 1-forms
$a_{I_{1}\cdots I_{n}}$ its supersymmetric partners.
The expansions of $E_{I}$ and of the ghost superfield $C$ read
\eq\ba{l}
E_{I}=e_{I}(x)+\dsum{n=1}{{\NT}} \frac{1}{n!}\theta
^{I_{1}}\cdots \theta ^{I_{n}}e_{II_{1}\cdots I_{n}}(x)\ ,\es
C=c(x)+\dsum{n=1}{{\NT}}\frac{1}{n!}\theta
^{I_{1}}\cdots \theta ^{I_{n}}c_{I_{1}\cdots I_{n}}(x)\ .
\ea\eqn{expansions-E-C}
The infinitesimal supergauge transformations of the superconnection
are expressed as the nilpotent BRST transformations
\begin{equation}
\cas \hA=-\hd C-[C,\hA]\ ,\quad \cas C=-C^{2}\ ,\quad \cas ^2=0\ .
\label{BRST-Transform}
\end{equation}
In terms of component superfields we have
\begin{equation}
\cas A=-dC-[C,A]\ ,\quad \cas E_{I}=-\partial _{I}C-[C,E_{I}]\ ,
\quad \cas C=-C^{2}\ .
\label{BRST-comp}\end{equation}
The supercurvature
\eq
\hF=\hd\hA + \hA^2 = F_A  + \Psi_I \;d\theta^I
+ \F_{IJ}\;d\theta^Id\theta^J
\eqn{superFStrenght}
transforms covariantly:
\[
\cas \hF = -[C,\hF]\ ,
\]
as well as its components 
\eq  
F_A  = dA+A^{2}\ ,\quad \Psi_I = \partial_{I}A + D_A E_{I}\ ,
\quad \F_{IJ} = \half\lp \pa_{I}E_{J}+\pa_{J}E_{I} + [E_{I},E_{J}] \rp\ ,
\eqn{superComp}
where the covariant derivative with repect to the connection
 $A$ is defined by
$D_A(\cdot )=d(\cdot )+[A,(\cdot )] $

Special cases, $\NT =1, 2$ and a discussion of the WZ-gauge can 
be found in \cite{nos}.

\section{Observables as equivariant cocycles}
\label{observables-equiv}

As discussed in~\cite{nos} and recalled in Appendix~\ref{app-d}, 
it is possible to suppress all the supergauge
degrees of freedom except the usual one corresponding to the 
$\theta=0$ component $c$ of the superghost $C$. 
This is the so-called ``WZ-gauge fixing'', obtained by fixing to 
zero a set of field components as in \equ{condi-NWZ}.
In  the WZ-gauge the supersymmetry generators must be 
modified into new operators $\tQ_I$ \equ{def-q-tilde}
differing from the
previous ones, $Q_I$, by a field dependent gauge transformation. 
Accordingly, the algebra of the supercharges $\tQ_I$ closes up to 
field dependent gauge transformations as in \equ{s-agebra-WZ}.

A possible generalization  for any $\NT$ of Witten's equivariant 
cohomology~\cite{witten-donald} may be defined as 
follows, in the WZ-gauge: 

\noindent -- An ``equivariant cocycle'' is a
{\it gauge invariant} local field polynomial -- integrated or not --
obeying to the conditions
\eq
\tQ_I\D=0\ ,\quad I=1,\cdots,\NT\ ;
\end{equation}
\noindent -- A ``trivial cocycle'' is an  equivariant cocycle of the form
\eq
\D = \tQ_1\cdots\tQ_\NT \D'\ ,
\end{equation}
where $\D'$ is {\it gauge invariant}.

\noindent -- The ``$\NT$-equivariant cohomology'' is the set of equivalence
classes of equivariant cocycles corresponding to the equivalence
relation
\eq
\D_1\approx\D_2 \quad \Leftrightarrow \quad
\D_1-\D_2 \ \ \mbox{is trivial}\ .
\end{equation}
This suggests the following generalization of Witten's definition: 
\begin{definition}
An ``observable'' is an element of the $\NT$-equivariant cohomology.
\end{definition}
{\bf Remark.} In the superspace formalism, a SUSY invariant has
necessarily the form of a total SUSY variation 
$\D = Q_1\cdots Q_\NT \D'$, as stated in Corollary 1 of
Proposition \ref{Proposition-4}~\footnote{The cohomology defined here
should not be confused with that of the nilpotent operator $Q=\sum_I\e^IQ_I$
or $\tQ=\sum_I\e^I\tQ_I$,
where the $\e^I$ are commuting constant supersymmetry ghosts. The latter 
may be used in matters such as perturbative renormalization. 
See e.g.~\cite{mag-pi-wo,pi-sor} in the context of supersymmetric gauge
theories.}.

\section{Observables as supersymmetric BRST cocycles}
\label{observables-brst}

\subsection{Defining the problem}

Computing the equivariant cohomology defined in the
preceding section is presumably a difficult task. Instead of this, we
shall generalize to arbitrary $\NT$ the appoach made in~\cite{boldo}
for the case $\NT=1$, defining observables, in the superspace formalism,
as elements of the BRST cohomology in the space of the SUSY invariant 
space-time integrals of local field polynomials.
Our task is thus to  define and find global observables, of the form
\begin{equation}
^{K} \Delta_{(d)} = \int_{M_d}\;\;   ^K \om^0_d \label{Delta}
\ ,
\end{equation}
 integral of a $p$-form  on a manifold $M_d$ of dimension $d$. 
The labels of a form $\,^S\om_p^g$ are defined as follows. $S$ is a $\NT$ 
component vector, with components
$(s_1, s_2, \cdots, s_{\NT})$ equal to
 the  (nonnegative) SUSY numbers, $p$ is the  
(nonnegative) form 
degree and $g$ is the (nonnegative) ghost number.

 The  observable (\ref{Delta}) has  by definition
to satisfy the BRST cocycle condition:
  \begin{equation}
\cas\,\,^{K}\Delta
_{d}=0,\,\,\,\,\,\,\,\,\,\,\,\,\,\,\,\,\,\,^{K}\Delta
_{d}\neq \cas\,\,\,^{K}\Delta _{d}^{\prime }\ , \label{um}
\end{equation}
$^{K} \Delta_{(d)}$ and $^{K} \Delta_{(d)}'$ being submitted to the
SUSY constraints
\begin{equation} 
Q_{I}\,\,\,^{K}\Delta_{d}=0\ ,\quad
Q_{I}\,\,\,^{K}\Delta_{d}'=0\ ,\qquad
I=1,2,\cdots ,\NT \label{dois}
\end{equation}

\subsection{General solution of  the SUSY constraints}

 We shall solve  (\ref{dois}) for $\,^{K}\Delta_{d}$, the solution
for $\,^{K}\Delta_{d}'$ being analogous. From (\ref{dois}) we obtain the
following equations for the integrand defined in \equ{Delta}:
\begin{equation}
Q_{I}\,\,^{(k_{1},\cdots k_{I},\cdots k_{\NT})}\omega
_{d}^{0}+d\,\,^{(k_{1},\cdots ,k_{I}+1,\cdots k_{\NT})}\omega
_{d-1}^{0}=0,\,\,\,\,\,\,\,\,\,\,\,\,\,I=1,2,\cdots ,\NT  \label{Qdescida}
\end{equation}
Using Proposition 4 of Appendix \ref{app-b},  
we conclude that the general solution is given by
\begin{equation}
^{K}\omega _{d}^{0}=Q_{1}\cdots Q_{\NT}\,\,^{K-E}\omega
_{d}^{0}\,\,\,+\,\,\,d\,\,^{K}\varphi _{d-1}^{0} \ , 
\label{5}\end{equation}
where $E$ is the $\NT$ dimensional vector
$E=(1,1,\cdots ,1)$.
From the identity \equ{sfieldform} obeyed by any superfield and the fact that
a product of more than $\NT$ operadores $Q_I$ is identically vanishing, 
we see that
we can replace the form $\,^{K-E}\omega_{d}^{0}$ in \equ{5} by the
superfield form
\begin{equation}
\,^{K-E}\Omega_d^0(x,\theta)=\,\,\exp \{\theta
^{I}Q_{I}\}\,\,^{K-E}\omega _d^0(x) \ ,
\label{sfieldform1}\end{equation}
and thus  we can  write \equ{5} as
\begin{equation}
^{K}\omega _{d}^{0}(x)=Q_{1}\cdots Q_{{\NT}}\,\,\,\,^{K-E}
\Omega _{d}^{0}(x,\theta)+ d\,\,^{K}\vf_{d-1}^0\ ,  
\label{eq omega}\end{equation}
and write (\ref{Delta}) as  a superspace integral\footnote{See the definition 
given by \equ{sup-integ} in Appendix \ref{app-a}.}
\begin{equation}
^{K}\Delta _{d}=\superint \,^{K-E}\Omega^{0}_d(x,\theta) \ .
\label{eq-form-sup}\end{equation}

\subsection{General solution of the BRST cocycle condition}

The BRST invariance condition (\ref{um}) yields, for  
the integrand of \equ{Delta}:
\[ 
\cas\,^{H+E}\omega _{d}^{0}+d\,^{H+E}\omega
_{d-1}^{1}=0\ ,
\] 
 for some form $^{H+E}\omega _{d-1}^{1}$.
From our previous result (\ref{eq omega}), this can be rewritten as
\eq
\cas\,\,Q_{1}\cdots Q_{\NT}\,\,^{H}\Omega
_{d}^{0}+d\,\,^{H+E}\omega _{d-1}^{1} = 0\ .
\eqn{eq-de-descida}
For convenience we have redefined the SUSY numbers by putting 
$H$ = $(h_1,\cdots,h_{\NT})$ = $K-E$ = 
 $(k_1-1,\cdots,k_{\NT}-1)$.
Let us show that the second term in the latter equation
can also be written as a total SUSY variation. Applying $Q_I$ to this 
equation we obtain  
\[
d Q_I\,^{H+E}\om^1_{d-1} = 0\ ,\quad I=1,\cdots,\NT\ ,
\]
hence, due to the triviality of the cohomology of $d$ in the space 
of local field functionals~\cite{barnich}:
\[
Q_I\,^{H+E}\om^1_{d-1} = d(\cdots) \ ,\quad I=1,\cdots,\NT\ .
\]
Application of Proposition 4 of Appendix \ref{app-b} with 
$\d_i=Q_I$ and $\d_n=d$ then yields
\[
\,^{H+E}\om^1_{d-1} = Q_{1}\cdots Q_{{\NT}}\,
\,^{K-E}\Omega _{d-1}^{1}(x,\theta)+ d(\cdots)\ ,
\]
and \equ{eq-de-descida} takes the form
\eq
Q_{1}\cdots Q_{\NT} \lp\cas\,\,^{H}\Omega_{d}^{0}
+d\,\,^{H}\Omega _{d-1}^{1}\rp = 0\ .
\eqn{eq-de-descida'}
Now, application of Proposition 2 of Appendix \ref{app-b} gives
\eq
\cas\,\,^{H}\Omega_{d}^{0} + d\,\,^{H}\Omega _{d-1}^{1}
+ \sum_I Q_I\,\,^{H-E_I}\Omega_{d}^{1} = 0\ .\quad
\eqn{first-mult-eq}
We shall show now that we can generate from the latter equation a
complete set of ``multi-descent equations'' -- a generalization of
the notion of descent equations to the case of more than two antiderivative 
operators, which are here the operators 
$\cas$, $d$, $Q_1$, $\cdots$, $Q_\NT$.
In order to do it, it is convenient to work with  ``truncated
superforms''~\cite{boldo}, a special case of the ``truncated
extended forms'' introduced in Appendix \ref{app-c}. 
We define a truncated superform of degree $q$ by
\eq
\trO_q = \left[ \hO_q\right]\truncation\ ,
\eqn{def-trunc}
where $\hO_q$ is a $q$-superform as defined by \equ{s-form}, 
which we may write as\footnote{The second summation in \equ{def-trunc}
is performed over all nonnegative values of
the SUSY numbers $s_I$, constrained by $|S|=k$, where 
$|S|\equiv\sum_{I=1}^\NT s_I$.}
\eq
\hO_q = \sum_{k=0}^{q}\sum_S^{|S|=k}
\,\,^S\OM_{q-k}\,(d\theta)^S
\equiv \sum_{k=0}^{q}\sum_{s_1,\cdots,s_\NT}^{s_1+\cdots+s_\NT=k}
\,\,^{(s_1,\cdots,s_\NT)}\OM_{q-k}\, 
(d\theta_1)^{s_1}\cdots (d\theta_\NT)^{s_\NT}\ ,
\eqn{def-s-form'}
and where truncation, labeled by the exponent ``$\truncation$'', 
means discarding, in the expansion \equ{def-s-form'}, 
all the terms of degrees $q-k>d$, $s_I>h_I$ ($I=1,\cdots,I_\NT$). 
Explicitly\footnote{Recall that all numbers such as form degree, SUSY number 
and ghost number, are nonnegative. As a general convention, any term
which may appear with negative such numbers is understood to vanish.}:
\eq
\trO_q = \sum_{k={\rm Max}\{0,q-d\}}^{q}\,\,\sum_{S}^{S\le H,\,|S|=k}
\,\,^S\OM_{q-k}\,(d\theta)^S\ ,
\eqn{def-trunc,}
where $S\le H$ means $s_I\le h_I$, $\forall I$. 
  We shall denote by $\tre$ the space of truncated superforms defined
by \equ{def-trunc}, \equ{def-trunc,}.

The exterior derivative $\trd$ acting in the space $\tre$   
is defined, according to \equ{trunc-alg} (Appendix \ref{app-c}), as
 \eq\ba{l}
\trd\trO_q = \left[\hd\trO_q\right]\truncation\es
=  \dsum{k={\rm Max}\{0,q+1-d\}}{q} \dsum{S}{S\le H,\,|S|=k}
     d\,\,^S\OM_{q-k}\,(d\theta)^S
  + \dsum{I=1}{\NT}\lp \dsum{k={\rm Max}\{0,q-d\}}{q}
    \,\,\dsum{S}{S\le H-E_I,\,|S|=k}
     \pa_I\,\,^S\OM_{q-k}\,d\theta^I(d\theta)^S\rp .
\ea\eqn{def-d-tr} 
  As shown in Appendix \ref{app-c}, $\trd$ is nilpotent. 
Proposition \ref{Proposition-C}  may be restated as:
\begin{lemma}\label{lemma2.1}
The cohomology of $\trd$ in the space $\tre$
consists of the truncated superforms of maximal weights:
\eq
\trO_D = \,\,^H\OM_d (d\theta)^H  \ ,\quad D\equiv d+|H|\ .
\eqn{cohom-trd}
\end{lemma} 
One checks easily that \equ{first-mult-eq} may be written in terms of
truncated superforms:
\eq
\cas \trO^0_D + \trd \trO^1_{D-1}=0\ ,
\eqn{first-mult-eq'}
where $D= d+|H|$, and the upper index as usual denotes the ghost number.
The two truncated superforms appearing in this equation  are
\eq
\trO^0_D= \,\,^H\OM^0_d\ ,\quad\quad
\trO^1_{D-1} = \,\,^{H}\Omega _{d-1}^{1} (d\theta)^H
+  \dsum{I=1}{\NT} \,^{H-E_I}\Omega_{d}^{1} (d\theta)^{H-E_I}\ .
\eqn{first-2-tr-s-forms}
Aplying $\trd$ to \equ{first-mult-eq'}, we obtain
\[
\trd \cas \trO^1_{D-1} = 0\ ,
\]
which, due to the triviality of the cohomology of $\trd$, solves in
\[
\cas \trO^1_{D-1} + \trd \trO^2_{D-2}\ .
\]
Repeating the argument we 
finally obtain the $(\cas,\,\trd)$ descent equations
\eq
\cas \trO^g_{D-g} + \trd \trO^{g+1}_{D-g-1}\ ,\quad g = 0,\cdots,D\ .
\eqn{m-desc}
These are the ``multi-descent equations'' which characterize the observable
\equ{Delta}, of dimension $d$ and SUSY weight $H+E$. These equations read,
explicitly\footnote{For the case of one SUSY generator (see \cite{boldo}) 
they are called bi-descent equations.}:
\eq\ba{c}
\cas\,^{S}\Omega _{p}^{g} \,+\, d\,\,^{S}\Omega
_{p-1}^{g+1} \,+\, \dsum{I=1}{\NT}Q_{I}\,^{S-E_{I}}\,\Omega
_{p}^{g+1}=0\ ,\es
g=D-p-|S|\ ,\quad
s_{I}=0,\cdots,h_{I}\ ,\quad p=0,\cdots ,d\ ,\quad
\ea\eqn{m-desc-explicit}
where $S=(s_{1},\cdots ,s_{\NT})$,  $H=(h_{1},\cdots ,h_{\NT})$,
$|S|=\sum_1^\NT s_I$,  $|H|=\sum_1^\NT h_I$ and $D=d+|H|$.

\subsection{Solving the multi-descent equations}
\subsubsection{Cohomology of $\cas$}

  With the purpose of resolving the multi-descent equations \equ{m-desc},
our first task will be to resolve the cohomology of the BRST operator 
$\cas$ in the space $\EE_S$ of the local polynomials in the superfield
forms of the theory and of their derivatives, and then in the space $\tre$
of the truncated superforms \equ{def-trunc}.  
An obvious  algebraic basis of $\EE_S$ 
is given by the set of 
superfields (\ref{expansion-A}-\ref{expansions-E-C}) 
and their derivatives:
\eq
\lac A,\,E_{I},\,C,\,A_{,\,I_1\cdots I_n},\,E_{I,\,I_1\cdots I_n},\,
C_{,\,I_1\cdots I_n},\,
dA,\,dE_{I},\,dC,\,dA_{,\,I_1\cdots I_n},\,dE_{I,\,I_1\cdots I_n},\,
dC_{,\,I_1\cdots I_n}   \,;\,n\ge1\rac \ ,
\eqn{basis1}
where  we use the notation 
$ X_{,\,I_1\cdots I_n} = \pa_{I_1}\cdots \pa_{I_n}X$
for the $\theta$-derivatives.

A more convenient basis is one that consists of BRST doublet superfields
and of covariant superfields. The BRST doublets are identified as
\eq
\lp E^{\rm (A)}_{I_1\cdots I_n},\,K^{\rm (A)}_{I_1\cdots I_n}\rp\ ,\quad
n\ge1\ ,
\eqn{brst-doublets}
where the $E^{\rm (A)}$ are the completely antisymmetrized 
$\theta$-derivatives of
$E_I$, and the $K^{\rm (A)}$ are their BRST variations:
\[
E^{\rm (A)}_{I_1\cdots I_n} = E_{[I_1,\,I_2\cdots I_n]}\ ,\quad
K^{\rm (A)}_{I_1\cdots I_n} = \cas E^{\rm (A)}_{I_1\cdots I_n}\ , \quad
\cas K^{\rm (A)}_{I_1\cdots I_n} = 0\ .
\]
 The remainder of the basis is given by the set of covariant superfields
constructed in Appendix \ref{app-d} and shown in Eq. \equ{curvature-basis}.
A complete algebraic basis is thus provided by:
\eq\ba{c}
\lac E^{\rm (A)}_{I_1\cdots I_n},\,K^{\rm (A)}_{I_1\cdots I_n},\,
dE^{\rm (A)}_{I_1\cdots I_n},\,dK^{\rm (A)}_{I_1\cdots I_n}
\,;\,n\ge1\rac\es
\bigoplus\es
\lac A,\,C,\,F_A,\,dC,\,\Psi_{I_1\cdots I_n}^{(\rm A)}\,(n\ge1),\,
\F^{(M)}_{I_1\cdots I_n}\,(n\ge2),\,
D_A\Psi_{I_1\cdots I_n}^{(\rm A)}\,(n\ge1),\,
D_A\F^{(M)}_{I_1\cdots I_n}\,(n\ge2) \rac   
\ea\eqn{basis3}
A first obvious conclusion is that the fields 
$E^{\rm (A)}_{I_1\cdots I_n}$ and $K^{\rm (A)}_{I_1\cdots I_n}$ and
their derivatives do not contribute to the cohomology of $\cas$, since they
are BRST doublets~\cite{pi-sor}. Moreover, the 
fields $\Psi_{I_1\cdots I_n}^{(\rm A)}$ and 
$\F^{(M)}_{I_1\cdots I_n}$ can be viewed 
as ``matter'' fields, transforming in the adjoint
representation of the gauge group. Then, as a consequence of the general
resuls of~\cite{barnich}, we can conclude that the 
cohomology of $\cas$ in the space $\EE_S$ consists of the local
polynomials generated by the cocycles
\eq
\theta_r(C)\quad (\, r=1,\dots , \mbox{rank }G \, )\quad \mbox{and} 
\quad P^{\rm inv}(F ,\, \Psi_{I_1\cdots I_n}^{(\rm A)} ,\, 
\F^{(M)}_{I_1\cdots I_n} ,\, D_A \Psi_{I_1\cdots I_n}^{(\rm A)} ,\,
D_A \F^{(M)}_{I_1\cdots I_n} )
\ ,
\eqn{s-cohom-EE_S}
where $P^{\rm inv} (\cdots)$ 
is any gauge invariant polynomial of its arguments, and 
where $\theta_r$ is the ghost cocycle 
associated in a standard way~\cite{barnich} to the $r^{\rm th}$ Casimir
operator of the gauge group $G$ and given, 
as function of the superghost $C$, by
\eq\ba{l}
\theta_r(C) = (-1)^{{m_r}-1} \ \dfrac{{m_r}! \, ({m_r}-1)!}{g_r !}
\ \tr C^{g_r} \quad 
(\, g_r=2m_r-1 \,,\ r= 1,\dots, \mbox{rank }G  \, )\ ,
\ea\eqn{s-cohom}
where  the
index $r$ labels the $r^{\rm th}$ Casimir
operator of the structure group (gauge group) $G$, whose degree is
denoted by $m_r$.
  An obvious generalization of the results of~\cite{barnich} shows
that  the cocycles (\ref{s-cohom}) 
are related by superdescent equations involving 
superforms $\lc \hat\theta_r \rc_p^{g_r-p}$
of form degree $p\geq 0$ and ghost-number $g_r-p$:
\eq\ba{c}
\cas \lc \hat\theta_r \rc_p^{g_r-p} 
+ \hd \lc \hat\theta_r \rc_{p-1}^{g_r-p+1}
=0
\qquad (\, p=0,\dots, g_r 
\, ) 
\ ,\es
\mbox{with}\quad 
\lc \hat\theta_r \rc_0^{g_r} = \theta_r(C)
\quad {\rm and} \quad 
\hd \lc \hat\theta_r \rc^0_{g_r} = f_r(\hF)\ ,
\ea\eqn{desc-theta}
where 
\eq
f_r(\hF) = \tr \hF^{{m_r}} \quad
\ ( \, r= 1,\dots, \mbox{rank }G  \, )\ ,
\eqn{s-F-cocycle} 
$\hF=\hd\hA+\hA^2$ being the supercurvature \equ{superFStrenght}.
According to the last of equations \equ{desc-theta}, the 
``bottom'' superform 
$\lc \hat\theta_r \rc^0_{g_r}$ is the 
Chern-Simons superform of degree $g_r$ 
associated to the $r^{\rm th}$
Casimir operator.

A straightforward generalization 
of the result \equ{s-cohom-EE_S} from superfield forms to truncated 
superforms yields the following lemma:
\begin{lemma}\label{lemma2.2}
The cohomology of $\cas$ in the functional space $\tre$ is given by the
truncated forms whose nonvanishing coefficients are polynomials
in the superfield forms given in \equ{s-cohom-EE_S}.
\end{lemma}

\subsubsection{Cohomology of $\cas$ modulo $\trd$}

The resolution of the cohomology of $\cas$ modulo $\trd$ in the space
$\tre$ of truncated superforms, i.e. the resolution of the multi-descent
equations \equ{m-desc} has been done in Appendix A.4 of
~\cite{boldo} for the case
$\NT=1$. The computation, relying on the cohomologies of 
$\trd$  and $\cas$ in $\tre$
(Lemmas \ref{lemma2.1} and  \ref{lemma2.2}), applies as well to 
arbitrary $\NT$, thus leading to the proposition:
\begin{proposition}\label{main-proposition}
The general solution of the multi-descent equations \equ{m-desc}
corresponding to the observable \equ{Delta} is generated, 
at ghost-number zero,  by two
classes of solutions. The first one is given by the superfield
forms (recall that $H=K-E$)
\eq\ba{ll}
\mbox{\rm Solution of Type I:}\qquad&
\;^{H}\OM^0_d 
\, (d\theta )^{H} 
 =\lc \lc\hat\theta_{r_1}\rc^0_{g_{r_1}} 
f_{r_2}(\hF) \cdots f_{r_L}(\hF) \rc_{S=H,\,p=d} \es
& \mbox{with}  \ \  |H|+d =D\ ,\quad D = 2\, \dsum{i=1}{L} m_{r_i} -1 
\ , \quad L\ge1\ ,
\ea\eqn{gen-sol-1}
where the Chern-Simons superform $\lc\hat\theta_{r}\rc^0_{g_{r}}$
and supercurvature invariant $f_{r}(\hF)$ are defined
by the equations (\ref{s-cohom}) - (\ref{s-F-cocycle}).

The second class of solutions depends on the 
superfield forms $F$, $\Psi$
and $\F^{(M)}$ appearing in the cohomology of $\cas$ 
(See \equ{s-cohom-EE_S})
and it is given by 
\eq
\mbox{\rm Solution of Type II:}\qquad
\;^{H}\OM^0_d = 
\;^{H}\ZZ^0_d \,(\,F ,\, \Psi_{I_1\cdots I_n}^{(\rm A)} ,\, 
\F^{(M)}_{I_1\cdots I_n} ,\, D_A \Psi_{I_1\cdots I_n}^{(\rm A)} ,\,
D_A \F^{(M)}_{I_1\cdots I_n}\, )
\ ,
\eqn{gen-sol-2}
Here, 
$\;^{H}\ZZ^0_d$ is an arbitrary invariant polynomial of its
arguments, 
which has a form degree $d$ and SUSY-numbers given by $H$,
and which is nontrivial in the sense that 
\[
\;^{H}\ZZ ^0_d \not= d \;^H\Phi_{d-1}^0   + 
\dsum{I=1}{\NT} Q_I  \;^{H-E_I}\Phi_{d}^0\ .
\] 
\end{proposition}

\noindent  {\bf Remarks.}

\noindent1. As in the $\NT=1$ case~\cite{boldo}
the superspace  integral (see \equ{eq-form-sup})
of any solution of the type \equ{gen-sol-2}:
\[
^K\D_d = \superint \;^{K-E}\ZZ^0_d\ ,
\]
which belongs to the BRST cohomology in the space of the 
SUSY invariant BRST cocycles, is in fact trivial from the point of view
of the equivariant cohomology defined in Subsection \ref{observables-equiv}.
Indeed, being the space-time integral of a superfield form
\[
^K\om_d^0 = 
-\dfrac{1}{\NT!} Q_1\cdots Q_\NT\;^{K-E}\ZZ^0_d\ ,\quad \mbox{where}\quad
^{K-E}\ZZ^0_d\quad \mbox{is gauge invariant}\ ,
\]
it reduces in the WZ-gauge to an equivariantly trivial 
expression
\[
\int_{M_d}\;^K\om_d^0 = 
-\dfrac{1}{\NT!} \tQ_1\cdots \tQ_\NT\int_{M_d}\;^{K-E}z^0_d\ ,
\quad \mbox{where}\quad
^{K-E}z^0_d\quad \mbox{is gauge invariant}\ .
\]
This follows from the fact that the operators $Q_I$ and $\tQ_I$ coincide, 
when applied to gauge invariant expressions, as seen from \equ{def-q-tilde}.

\noindent2.  As we shall see in Section \ref{witten-observ}, 
solutions of the type \equ{gen-sol-1} are not trivial from 
the point of view of the equivariant cohomology. In the case 
$\NT=1$~\cite{boldo} they are the Witten-Donaldson 
observables~\cite{witten-donald}. The cases $\NT>1$ offer thus a
generalization of the latters.

\subsection{Superform expression of the observables}

Let us consider the untruncated version of the equations \equ{m-desc}
involving full superforms (see \equ{s-form}):
\eq
\cas \hO^g_{D-g} + \hd \hO^{g+1}_{D-g-1}\ ,\quad g = 0,\cdots,D\ 
\quad (D=|H|+d)\ .
\eqn{super-desc}
It can be shown~\cite{boldo} on the basis of the results 
of~\cite{barnich} about BRST cohomology, that the general solution of 
the super-descent equations
\equ{super-desc} which contains superforms down to and including 
ghost number $g=0$, is given by
\eq
\hO^{g_{r_1}-p}_{D-g_{r_1}+p} = \lc{\hat\theta}_{r_1} \rc^{g_{r_1}-p}_p
f_{r_2}(\hF)\cdots f_{r_L}(\hF)\ ,
\quad p=0,\dots,g_{r_1} \ ,  
\eqn{3.x}
with the $\lc{\hat\theta}_{r_1} \rc^{g_{r_1}-p}_p$
and the supercurvature invariant $f_{r}(\hF)$ defined
by the equations (\ref{s-cohom}) - (\ref{s-F-cocycle}).
The superfield components $\;^S\OM^g_p$, with$|S|+g+p=D$,
of these superforms are clearly solutions of the multi-descent descent
equations \equ{m-desc} since the latter
is a subsystem of \equ{super-desc}. 
The corresponding observables are given by the superspace integrals of
the superfield components of the expansion 
\eq
\hO^{0}_{D}= \lc{\hat\theta}_{r_1} \rc^{0}_{g_{r_1}}
f_{r_2}(\hF)\cdots f_{r_L}(\hF)
 \equiv \dsum{H}{|H|\le D} \;^{H}\Omega^0_{D-|H|} \,(d\theta)^{H}\ ,
\eqn{Omega0-D}
i.e. (see \equ{Delta} and \equ{eq-form-sup})
\eq
^K\D_d = \superint\;^H\OM^0_d\ ,\qquad d=D-|H|\ ,
   \quad K=H+(1,\cdots,1)\ .
\eqn{observable-K-D}
On the other hand, we see from the equations \equ{Omega0-D} 
taken with all possible values of $D$ and of the numbers $g_{r_i}$ 
that the components of the
solutions of the superdescent equations \equ{super-desc} span all the
solutions of type  I  \equ{gen-sol-1} of the multidescent equations \equ{m-desc}.
Thus:
\begin{proposition}\label{super-main-proposition}
Let $\hO^{0}_{D}$ represent the general solution of the 
super-descent equations \equ{super-desc}. Then 
the superfield forms $\;^{H}\Omega^0_{D-|H|}$ 
defined by the expansion of  $\hO^{0}_{D}$ in \equ{Omega0-D}
represent the general solution of type \equ{gen-sol-1}
of the multi-descent equations \equ{m-desc}.
\end{proposition}
We note for the sake of completeness
that the superforms $\hO^{g_{r_1}-p}_{D-g_{r_1}+p}$ 
obey the system of superdescent equations
\[
\cas \hO^{g_{r_1}-p}_{D-g_{r_1}+p}
 + \hd \hO^{g_{r_1}-p+1}_{D-g_{r_1}+p-1}\ ,\quad
p=0,\dots,g_{r_1} \ ,
\]
involving ghost numbers up to the value  $g_{r_1}$, which is less 
 than the maximum possible value $D$ if $L>2$.

A convenient way of representing the observables
\equ{observable-K-D} and deducing
interesting properties of them, is based on
the identity $\hd\lc{\hat\theta}_{r_1} \rc^{0}_{g_{r_1}}$ = 
$f_{r_1}(\hF)$ for the Chern-Simons form, and the expansion
\eq
\hd\hO^0_D = f_{r_1}(\hF)\cdots f_{r_L}(\hF)
= f_{r_1}(F)\cdots f_{r_L}(F) 
+\dsum{S}{1\le|S|\le D+1}\;^SW^0_{D+1-|S|} (d\theta)^S\ ,
\eqn{f-f-f}
with the first term being a $d$-derivative, and
\eq
^SW^0_{D+1-|S|} = \dsum{I=1}{\NT} Q_I\;^{S-E_I}\Omega^0_{D+1-|S|}
  + d \;^{S}\Omega^0_{D-|S|}.
\eqn{W=Q+d}
Integrating both sides of the latter equation in superspace, 
we see that we can write \equ{observable-K-D} as
\eq\ba{l}
^K\D_d = \superint\;^H\OM^0_d 
= (-1)^{\NT-J}\dint_{M_d} Q_1\cdots\widehat{Q_J}\cdots Q_\NT
  \;^{H+E_J}W^0_{d}\ ,\es
\mbox{with}\quad H=K-E\ ,\quad  d = D-|H|= D+\NT-K\ ,
\ea\eqn{obser-W}
where the notation $\widehat X$ means suppression of the factor $X$.
The value of $J$ in the right-hand side is arbitrary. Let us show that the  
expression is in fact independent of $J$ as it should.
Applying the nilpotent operator $\hd$ on \equ{f-f-f} we obtain the
descent equations
\eq\ba{l}
\dsum{I=1}{\NT} Q_I \;^{H-E_I}W^0_{D+2-|H|} + d\;^{H}W^0_{D+1-|H|}= 0\ ,
\quad 1\le|H|\le D+1\ ,\es
\dsum{I=1}{\NT} Q_I \;^{H-E_I}W^0_{0}=0\ , \quad |H| = D+2\ .
\ea\eqn{W-descent-eq}
Considering the difference of the expressions \equ{obser-W} obtained for
two values of $J$, which we may choose without loss of generality as
$J$ = $\NT$ and $\NT-1$, respectively, we obtain
\[
\dint_{M_d} Q_1\cdots Q_{\NT-2}
\lp  Q_{\NT-1} \;^{H'-E_{\NT-1}}W^0_{d}
  + Q_{\NT} \;^{H'-E_{\NT}}W^0_{d} \rp\ ,\quad H'=H+E_{\NT}+E_{\NT-1}\ ,
\]
which, by virtue of \equ{W-descent-eq}  for $|H|=D+2-d$, reads
\[
 -\dint_{M_d} Q_1\cdots Q_{\NT-2}\,
\dsum{I=1}{\NT-2}  Q_{I} \;^{H'-E_I}W^0_{d} = 0 \ ,
\]
and which vanishes due to the nilpotency of the operators $Q_I$.

\section{Witten's observables and descent 
equations}\label{witten-observ}

Let us rewrite the integral \equ{obser-W}, expressing a generic observable, as a 
space-time integral:
\eq
^K\D_d = \dint_{M_d} \;^K\om_d\ ,
\eqn{obs-space-int}
the integrant being defined up to a total space-time
derivative. Let us define the latter as
\eq
\left. \;^K\om_d = \dsum{J=1}{\NT} (-1)^{\NT-J} \a_J\, 
 Q_1\cdots\widehat{Q_J}\cdots Q_\NT \;^{H+E_J}W^0_{d}\right|_{\te=0}
\ ,\quad \mbox{with}\quad \sum_J\a_J=1\ .
\eqn{alfa-def-integrant}
It is clear from the discussion at the end of last 
Subsection that, to the contrary of its integrant, the integral does
not depend on the arbitrary numbers $\a$. Calculating
\[
Q_I\;^K\om_d = (-1)^{\NT-1} \a_I\, Q_1\cdots Q_{\NT} \;^{H+E_I}W^0_{d}\ ,
\]
and
\[\ba{lll}
d \;^{K+E_I}\om_{d-1} &=& \left. \dsum{J=1}{\NT} (-1)^{J-1} \a_J\, 
 Q_1\cdots\widehat{Q_J}\cdots Q_\NT \,d \;^{H+E_J+E_I}W^0_{d-1}\right|_{\te=0}\es
&=& (-1)^{\NT} \a_I\, Q_1\cdots Q_{\NT} \;^{H+E_I}W^0_{d}\ ,
\ea\]
where we have used \equ{W-descent-eq} for the last equality, we
conclude that the integrants \equ{alfa-def-integrant} obey the descent
equations
\eq
Q_I\;^K\om_d  + \a_I\, d \;^{K+E_I}\om_{d-1} = 0 \ ,\quad
I=1,\cdots,\NT\ ,\quad \NT\le |K| \le D-d+\NT\ .
\eqn{Q-Witten-desc}
We can now go to the WZ-gauge (see Appendix \ref{app-d}). The forms
$\;^K\om_d$ being gauge invariant functions of the covariant superfields
\equ{curvature-basis} taken at $\te=0$, they
reduce to correspondent gauge invariant
functions of the covariant WZ-gauge fields \equ{cov-WZ-basis} by virtue of the 
correspondence \equ{bijection}. Moreover, since the expressions are
gauge invariant, the applications of generators $\tQ_I$ and $Q_I$ are
identical. Hence, the equations \equ{Q-Witten-desc} reduce to
\eq
\tQ_I\;^K\om_d  + \a_I\, d \;^{K+E_I}\om_{d-1} = 0 \ ,\quad
I=1,\cdots,\NT\ ,\quad \NT\le |K| \le D-d+\NT\ ,
\eqn{tilde-Q-Witten-desc}
which are the possible generalizations to arbitrary $\NT$ of Witten's
descent equations~\cite{witten-donald}. 

Specializing to two particular
values of the set of numbers $\a_I$, we would obtain:
\eq\ba{ll}
\tQ_I\;^K\om_d  + \dfrac{1}{\NT}\, d \;^{K+E_I}\om_{d-1} = 0 \quad 
&(\a_I=\dfrac{1}{\NT})\ ,\es
  \left.\ba{l} 
  \tQ_J\;^K\om_d  + \, d \;^{K+E_J}\om_{d-1} = 0 \es
  \tQ_I\;^K\om_d = 0 \ ,\quad
  I\not=J
  \ea \right\}
\quad &(\a_J=1\,,\;\,\a_I=0\,,\;I\not=J)\ .
\ea\eqn{special-Witten-desc}
Let us recall that in all systems of equations above, any term with
negative SUSY number or form degree is assumed to vanish.

The equations \equ{tilde-Q-Witten-desc} show that our 
solutions, which solve Witten's descent equations,
are indeed Witten's observables, their space-time integrals 
being $\tQ_I$-invariant
for any $I$. It remains to show that they are nontrivial in the sense 
defined in Section \ref{observables-equiv}. For this it is sufficient to
check the nontriviality of the $w$'s of highest SUSY numbers 
(hence of zero form degree). The latters are $\tQ_I$ invariant, and read (see
e.g. \equ{alfa-def-integrant} for $\a_1=1$, $\a_I=0\,,I\not=1$,  
and \equ{f-f-f})
\[
^{I_1J_1 \cdots I_nJ_n}w_0 = (-1)^{\NT-1} Q_2\cdots Q_\NT 
\left.\lp \F^{\phantom{\rm M}}_{I_1J_1}, \cdots, \F_{I_nJ_n} \rp
\right|_{{\rm symmetrized\ in\ }(I_1,\cdots, J_n)} \ ,\quad
\mbox{with}\quad n=\sum_{r=1}^L m_r\ ,
\]
where we use the notation \equ{def-s-form'} and
$(X_1,\cdots,X_n)$ is a symmetric invariant polynomial of its arguments.
In the WZ-gauge:
\[
^{I_1J_1 \cdots I_nJ_n}w_0 = (-1)^{\NT-1}\tQ_2\cdots \tQ_\NT 
\left.\lp e^{\rm M}_{I_1J_1}, \cdots, e^{\rm M}_{I_nJ_n} \rp
\right|_{{\rm symmetrized\ in\ }(I_1,\cdots, J_n)} \ ,\quad
\mbox{with}\quad n=\sum_{r=1}^L m_r\ .
\]
It is easy to check that the $e^{\rm M}_{IJ}$ can never be written as 
a $\tQ_1$-variation.
Hence $w_0$ cannot be written as a full 
$\tQ_1\tQ_2\cdots \tQ_\NT$-variation and thus belongs to the
equivariant cohomology.

\section{Conclusion and open problems}\label{conclusion}

The results on the classification of the observables known for the
topological Yang-Mills theories with one supersymmetry generator, 
were generalized to the case of theories defined with more
supersymmetry generators. We have obtained a complete
classification of the observables according to their definition as
nontrivial BRST cocycles in the space of supersymmetry invariant local
functionals. 

Although our solutions are solutions of the 
equivariant cohomology problem defined in Section \ref{observables-equiv},
we have no proof that it provides the complete solution of the latter.
However, we have also found generalized Witten's descent equations
for the integrants of the observables and showed 
their nontriviality in the equivariant cohomology sense.

Our results are
formal, being established in the classical approximation. Their
interpretation at the quantum level as topological invariants remains an  
open problem in the general case of arbitrary numbers of SUSY generators
and space-time dimensions, although some results are known 
for special cases, in particular for 
$\NT=2$~\cite{blau-thom-96,geyer}.

\vspace{5mm}

\noindent{\bf Acknowledgements.} We are much indebted to Jos\'e Luiz
Boldo for his collaboration at a preliminary stage of this work, and to
Jos\'e Alexandre Nogueira and Wander G. Ney
for many useful discussions.  C.P. Constantinidis would like to thank the
Abdus Salam International Center of Theoretical Physics (ICTP) for
hospitality during a visit under the Associate Program.

\newpage]

\appendix

\setcounter{section}{0}


\section*{Appendices} 
\label{notations}

\section{$\NT$- supersymmetry and superspace}\label{app-a}

$(D,\NT)$-superspace bosonic coordinates are denoted by $x^\m$,
$\m=0,\dots,D-1$, the fermionic (Grassmann, or anticommuting)
coordinates being denoted by
$\te^I$, $I-1,\dots,\NT$.
The $\NT$ supersymmetry generators $Q_I$ are represented on superfields
$F(x,\te)$ by
\[
Q_I F = \pa_I F \equiv \dpad{}{\te^I}F\ ,
\]
where, by definition, $\pa_K\te^J=\d_K^J$. Further conventions
 and properties about the
$\te$-coordinates are the following:
\[\ba{l}
\te^{\NT} = \e_{I_1\cdots I_{\NT}}\te^{I_1}\cdots\te^{I_{\NT}}
= \NT!\; \te^1\cdots\te^{\NT}  \ ,\es
(\dth)^{\NT} =\e^{I_1\cdots I_{\NT}}\pa_{I_1}\cdots\pa_{I_{\NT}}
= \NT!\; \pa_1\cdots\pa_{\NT} \ ,\es
(\dth)^{\NT}\te^{\NT} = - (\NT!)^2\ ,

\ea\]
where $\e^{I_1\cdots I_{\NT}}$
is the completely antisymmetric tensor of rank $\NT$, with
the conventions
\[
\e^{1\cdots \NT }=1\ ,\quad
\e_{I_1\cdots I_{\NT}} = (-1)^{\NT+1} \; \e^{I_1\cdots I_{\NT}}\ .
\]
One may define the conserved supersymmetry number -- {\it SUSY number} --
attributing the value 1 to the generators $Q_I$, hence $-1$ to the
$\te$-cordinates. The SUSY number of each field
component is then deduced from the
SUSY number given to each superfield.

Superspace integration of a superfield form $\OM_p(x,\te)$ is defined
by integrals
\eq 
\superint  \OM_p(x,\te) = \dint_{\!\!\!\!\!\!M_p}\;\dint\dnth \OM_p(x,\te)\ ,
\eqn{sup-integ}
where the $x$-space integral is made on some $p$-dimensional
(sub)manifold $M_p$, and the $\te$-space integral is the Berezin
integral defined by
\[
\dint\dnth\cdots = - \dfrac{1}{(\NT!)^2}(\dth)^{\NT}\cdots \ ,\quad
\mbox{such that}\quad \dint\dnth \te^{\NT}=1\ .
\]

\section{Some useful propositions}\label{app-b}

The propositions and proofs presented here are   generalizations of 
results given in~\cite{barnich}.  They hold for both usual forms 
and superfield forms.

\noindent
{\bf Definitions and notations.} 
Let $\omega ^{(s_{1},\cdots ,s_{n})}$ be forms
whose weights  $s_{i}$ are associated to $n$ operators $\delta _{i},$
$(i=1,\cdots ,n)$, nilpotent and anticommuting, \ie $\{\delta
_{i},\delta _{j}\}=0$. The cohomology group of each operator
  $\delta _{i}$ is  trivial by hypothesis. If some of the weights $s_{i}$
are negative we have, by 
convention,
  $\omega ^{(s_{1},\cdots ,s_{n})}=0$.
We shall use the condensed notations
\eq\ba{l}
\omega ^{(s_{1},\cdots ,s_{n})}=\omega^S\ ,\quad
S=(s_1,\cdots,s_n)\ ,\quad |S|=\dsum{i=1}{n}s_i\ ,\quad \es
E_i =  (0,\cdots,1,\cdots,0)\quad\mbox{(unique nonvanishing component 
is a 1  at the $i^{\rm th}$ position)}\ ,\es
E=\dsum{i=1}{n}E_i = (1,1,\cdots,1)\ ,\es
S-T=(s_1-t_1,\cdots,s_n-t_n)\ ,\quad
S\le T\ \Leftrightarrow\ s_i\le t_i\,,\ i=1,\dots n\ .
\ea\eqn{multi-indices} 
  The forms $\om^S$ may be fields or superfields.
\begin{proposition}\label{Proposition-1}
Let the set of forms 
$\{\omega ^{T-E_i}\,|\,i=1,\cdots ,n\} = 
\{\omega ^{(t_{1},\cdots ,t_{i}-1,\cdots ,t_{n})}\,|\,i=1,\cdots ,n\}$
satisfy the cocycle condition
\begin{equation}
\sum_{i=1}^{n}\delta _{i}\,\,\omega
^{T-E_i}=0\ . 
\label{7}\end{equation}

\noindent
{\bf 1.} The set $\{\omega ^{T-E_i}\,|\,i=1,\cdots ,n\}$
can be extended to an
{\it extended form} $\tilde\om$, defined by
\begin{equation}
\tilde{\omega}=\dsum{S}{|S|=|T|-1}\omega^{S}\ ,  
\label{8}\end{equation}
such that
\begin{equation}
\tilde{\delta}\,\,\tilde{\omega}=0\;;\,\,\,\,\,\,%
\,\,\,\,\,\,\tilde{\delta}=\sum_{i=1}^{n}\delta _{i}\,.  \label{9}
\end{equation}

\noindent
{\bf 2.} There exists an extended form
\begin{equation}
\tilde{\varphi}=\dsum{S}{|S|=|T|-2}\varphi^{S},  
\label{10}
\end{equation}
where  $\tilde{\omega}$ and $\tilde{\varphi}$ satisfy
\begin{equation}
\tilde{\omega}=\tilde{\delta}\,\,\,\tilde{\varphi},  \label{11}
\end{equation}
\end{proposition}
\begin{corollary}\label{Corollary-1-1} 
The cohomology of $\tilde{\delta}$\
is trivial. 
\end{corollary}
\begin{corollary}\label{Corollary-1-2} The general solution of (\ref{7})\
for any $\omega ^{T-E_i}$ is given by
\begin{equation}
\omega ^{T-E_i}=\sum_{j=1}^{n}\delta_{j}
\varphi^{T-E_i-E_j}\ ,\quad j=1,\cdots ,n\ , 
\end{equation}
where all $\varphi ^{T-E_i-E_j}$\ for $i,\,j=1,\cdots ,n$
are components of a single extended form
$\tilde{\varphi}$.
\end{corollary}
{\bf Proof of Proposition \ref{Proposition-1}. }  
The proof will be performed by induction from the case $n=2$
which will be first treated explicitly.

\noindent
{\bf Case $n=2$.} In this case, equation (\ref{7})
is given by
\begin{equation}
\delta _{1}\omega ^{(t_{1}-1,t_{2})}+\delta _{2}\omega
^{(t_{1},t_{2}-1)}=0\ . 
\label{eq-n=2}
\end{equation}

\noindent
{\it Proof of Part 1.} Applying $\delta_{1}$ to (\ref{eq-n=2})\
 we obtain: $\delta_{2}\delta_{1}\omega
^{(t_{1},t_{2}-1)}=0$. Remembering that the cohomology of $\delta _{2}$
is trivial, we deduce the existence of a form $\omega
^{(t_{1}+1,t_{2}-2)},$ such that:
\begin{equation}
\delta _{1}\omega ^{(t_{1},t_{2}-1)}+\delta _{2}\omega
^{(t_{1}+1,t_{2}-2)}=0 \ .
\label{13}
\end{equation}
 Repeating  successively this procedure we finally get
\begin{equation}
\delta _{1}\omega ^{(|T|-1,0)}=0\ ,\quad |T|=t_1+t_2\ . 
\label{14}\end{equation}
We have thus obtained the set of equations
\[
\delta _{1}\omega ^{(t_{1}+k-1,t_{2}-k)}+\delta _{2}\omega
^{(t_{1}+k,t_{2}-k-1)}=0\ ,\quad 0\leq k\leq t_{2}\ .
\]
Applying now  $\delta _{2}$\ to (\ref{eq-n=2}) and using the triviality of the cohomology of
 $\delta _{1}$, we obtain in an analogous way the following equations
\[
\delta _{1}\omega ^{(t_{1}-k^{\prime }-1,t_{2}+k^{\prime
})}+\delta _{2}\omega ^{(t_{1}-k^{\prime },t_{2}+k^{\prime
}-1)}=0\ ,\quad 0\leq k^{\prime }\leq t_{1}\ .
\]
These last two systems of equations can be put into a unique set:
\begin{equation}
\delta _{1}\omega ^{(t_{1}+p-1,t_{2}-p)}+\delta _{2}\omega
^{(t_{1}+p,t_{2}-p-1)}=0\ ,\quad -t_{1}\leq p\leq t_{2}\ ,
\label{component}
\end{equation}
which is exactly (\ref{9}) written in components, 
corresponding to the extended form
and to the extended operator
\begin{equation}
\tilde{\omega}=\sum_{p=-t_{1}}^{t_{2}}\omega^{(t_{1}+p-1,t_{2}-p)}\ ,\quad
\tilde{\delta}=\delta_{1} + \delta_{2}\ .
\end{equation}

\noindent
{\it Proof of part 2.} Equation (\ref{component}) for
$p=-t_{1}$
is (\ref{14}). From the triviality of the cohomology of
$\delta _{2}$ we obtain the general solution
\begin{equation}
\omega ^{(0,|T|-1)}=\delta _{2}\varphi ^{(0,|T|-2)}\ ,  
\label{17}
\end{equation}
and by substituting (\ref{17})  in  (\ref{component}) for $p=-t_{1}+1$,
we obtain
 \begin{equation}
 \delta _{2}\left[ -\delta _{1}\varphi ^{(0,|T|-2)}+\omega
^{(1,|T|-2)}\right] =0\ ,
\end{equation}
whose general solution for
$\omega^{(1,|T|-2)}$ is
\begin{equation}
\omega ^{(1,|T|-2)}=\delta _{1}\varphi ^{(0,|T|-2)}+\delta _{2}\varphi
^{(1,|T|-3)}\ .
\end{equation}
The procedure continues until
equation (\ref{component}) for  $p=t_{2}-1$, leading finally to
\begin{equation}
\omega ^{(t_{1}+p-1,t_{2}-p)}=\delta _{1}\varphi
^{(t_{1}+p-2,t_{2}-p)}+\delta _{2}\varphi
^{(t_{1}+p-1,t_{2}-p-1)}\ ,\quad -t_{1}+1\leq p\leq t_{2}\ . 
\label{19}
\end{equation}
with the last equation, for
$p=t_{2}$,
being identically satisfied. Notice that the set
(\ref{19}) can be put in the form
(\ref{11}) with
\begin{equation}
\tilde{\varphi}=\sum_{p=-t_{1}+2}^{t_{2}}\varphi
^{(t_{1}+p-2,\,t_{2}+p)}\ .
\end{equation}

\subsubsection*{\bf The general case}

In order to establish the proof for general $n$ we suppose that  the propositon is valid for $(n-1)$.
Applying for example  $\delta _{1}$ on  (\ref{7}), we get
\begin{equation}
\sum_{i=2}^{n}\delta _{i}\delta _{1}\omega
^{(t_{1},t_{2},\cdots ,t_{i}-1,\cdots ,t_{n})}=0\ .
\end{equation}
By the induction hypothesis there exists $(n-1)$ forms
$\omega
^{(t_{1}+1,t_{2}-2,t_{3},\cdots ,t_{n})}$ and \\ 
$\omega
^{(t_{1}+1,t_{2}-1,t_{3},\cdots ,t_{i}-1,\cdots ,t_{n})},$ with
$i=3,\cdots ,n,$ such that
\begin{equation}
\delta _{1}\omega ^{(t_{1},t_{2}-1,t_{3},\cdots ,t_{n})}+\delta
_{2}\omega
^{(t_{1}+1,t_{2}-2,t_{3},\cdots ,t_{n})}+\sum_{i=3}^{n}\delta
_{i}\omega ^{(t_{1}+1,t_{2}-1,t_{3},\cdots ,t_{i}-1,\cdots ,t_{n})}=0\ .
\label{20}
\end{equation}
Repeating the procedure we get
\begin{equation}
\delta _{1}\omega ^{(t_{1}+k-1,t_{2}-k,t_{3},\cdots ,t_{n})}+\delta
_{2}\omega
^{(t_{1}+k,t_{2}-k-1,t_{3},\cdots ,t_{n})}+\sum_{i=3}^{n}\delta
_{i}\omega
^{(t_{1}+k,t_{2}-k,t_{3},\cdots ,t_{i}-1,\cdots ,t_{n})}=0\ ,
\end{equation}
with $0\leq k\leq t_{2}$.
Beginning the same procedure from (\ref{7}), but applying $\delta _{2}$
and considering the cohomology of $\delta _{1}$, $\d_3,\cdots,\d_n$,                        
we obtain
\begin{equation}\ba{l}
\delta _{1}\omega ^{(t_{1}-k^{\prime }-1,t_{2}+k^{\prime
},t_{3},\cdots ,t_{n})}+\delta _{2}\omega ^{(t_{1}-k^{\prime
},t_{2}+k^{\prime }-1,t_{3},\cdots ,t_{n})}+\sum_{i=3}^{n}\delta
_{i}\omega ^{(t_{1}-k^{\prime
},t_{2}+k^{\prime },t_{3},\cdots ,t_{i}-1,\cdots ,t_{n})}=0\ ,\es
0\leq k^{\prime }\leq t_{1}\ .
\ea\end{equation}
We can unify the last two set of equations in the following one
\eq\ba{l} 
\delta _{1}\omega ^{(t_{1}+p-1,t_{2}-p,t_{3},\cdots ,t_{n})}+\delta
_{2}\omega
^{(t_{1}+p,t_{2}-p-1,t_{3},\cdots ,t_{n})}+\sum_{i=3}^{n}\delta
_{i}\omega
^{(t_{1}+p,t_{2}-p,t_{3},\cdots ,t_{i}-1,\cdots ,t_{n})}=0\ ,\es
-t_{1}\leq p\leq t_{2}
\ea\eqn{eqgeral} 
Introducing now the $2-$extended operator
$\tilde{\delta}_{(1,2)}=\delta _{1}+\delta _{2}$ and the 
sets of $2-$extended forms
 \eq\ba{l} 
\tilde{\omega}_{(1,2)}^{(\tilde{t},t_{3},\cdots ,t_{i}-1,\cdots ,t_{n})}
= 
\dsum{s_1,s_2}{s_{1}+s_{2} =\tilde{t}}
\omega ^{(s_{1},s_{2},t_{3},\cdots ,t_{i}-1,\cdots t_{n})}\ ,
\quad i=3,\cdots,n\ ,\es
\tilde{\omega}_{(1,2)}^{(\tilde{h}-1,t_{3},\cdots ,t_{n})}
= 
\dsum{s_1,s_2}{s_{1}+s_{2} =\tilde{t}-1}
\omega ^{(s_{1},s_{2},t_{3},\cdots t_{n})}\ ,
\ea\eqn{22}
we can rewrite  (\ref{eqgeral}) as:
\begin{equation}
\tilde{\delta}_{(1,2)}
\tilde{\omega}_{(1,2)}^{(\tilde{t}-1,t_{3},\cdots ,t_{n})}+%
\sum_{i=3}^{n}\delta _{i}
\tilde{\omega}_{(1,2)}^{(\tilde{t},t_{3},\cdots ,t_{i}-1,\cdots ,t_{n})}
=0\ ,\quad \tilde{t}=t_{1}+t_{2} \ .  
\label{eqestendida}
\end{equation}

Observe that the operator  $\tilde{\delta}_{(1,2)}$ is nilpotent and
anticommutes with the others $\delta _{i}$. 
By virtue of Proposition \ref{Proposition-1} already 
proved for the case $n=2$,
its cohomology is trivial. In order to solve
(\ref{eqestendida}) we make use of  Proposition \ref{Proposition-1}, 
true for $(n-1),\,$ by assumption, where the $(n-1)\,$
operators are given by $\{\tilde{\delta}_{(1,2)},\delta
_{3},\cdots ,\delta _{n}\}$. We have thus the extended form
\begin{equation}
\dsum{\tilde{s},s_3,\cdots,s_n}
{\tilde{s}+s_{3}+\cdots+s_n=|T|-1}\tilde{\omega}%
_{(1,2)}^{(\tilde{s},s_{3},\cdots ,s_{n})}=
\dsum{s_1,\cdots,s_n}
{s_{1}+\cdots+s_n=|T|-1}{\omega}^{(s_{1},\cdots ,s_{n})}
\,\equiv\, \tilde{\omega}\ ,
\end{equation}
satisfying (\ref{7}):
\[
\left( \tilde{\delta}_{(1,2)}+\sum_{i=3}^{n}\delta _{i}\right) \tilde{\omega}
= \,\tilde{\delta}\,\tilde{\omega}=0\ ,\quad
\tilde{\delta}=\sum_{i=1}^{n}\delta _{i}\ .
\]
We also know from {\it part 2} of Proposition \ref{Proposition-1} 
for $(n-1)$ 
that there exists an extended form
\[
\tilde{\varphi}=\dsum{\tilde{s},s_3,\cdots,s_n}
{\tilde{s}+s_{3}+\cdots+s_n=|T|-2}
\tilde{\varphi}_{(1,2)}^{(\tilde{s},s_{3},\cdots ,s_{n})} =
\dsum{s_1,\cdots,s_n}{s_{1}+\cdots+s_n=|T|-2}
{\varphi}^{s_1,\cdots,s_n}\equiv\tilde{\vf}\ ,
\]
which satisfies
\begin{equation}
\tilde{\omega}=\left( \tilde{\delta}_{(1,2)}+\sum_{i=3}^{n}\delta
_{i}\right) \,\tilde{\varphi}\equiv
\tilde{\delta}\,\,\tilde{\varphi}\ . 
\end{equation}
\qed

\begin{proposition}\label{Proposition-2}
If the form $\omega = \om^T =\omega^{(t_{1},\,\,\cdots \,\,,t_{n})}$ 
satisfies the equation
\begin{equation}
\delta _{1}\cdots \delta _{n}\,\omega =0.  \label{24}
\end{equation}
it admits a solution of the type
\begin{equation}
\omega =\sum_{i=1}^{n}\delta _{i}\,\varphi
^{T-E_i}. \label{25}
\end{equation}
\end{proposition}
{\bf Proof of Proposition \ref{Proposition-2}.} 
The proof is by induction.
For the case $n=1$ equation (\ref{24}) reads
$\delta \,\omega =0$, and from the triviality of the cohomology of
 $\delta $ the solution is given by
\[
\omega =\delta \,\varphi\ .
\]
For the general case, we can rewrite (\ref{24}) as
\begin{equation}
(\delta _{1}\cdots \delta _{n-1})\delta _{n}\,\omega =0\ ,
\label{prv2}
\end{equation}
and supposing that Proposition \ref{Proposition-2} 
is valid for ($n-1$), we can solve
(\ref {prv2}) with respect to $\delta _{n}\omega$,
obtaining
\begin{equation}
\delta _{n}\,\,\omega
^{(t_{1},\cdots ,t_{n}-1)}=\sum_{i=1}^{n-1}\delta _{i}\eta
^{T-E_i}\ .
\end{equation}
From Proposition \ref{Proposition-1} it follows that
\begin{equation}
\omega =\sum_{i=1}^{n}\delta _{i}\,\varphi
^{T-E_i}  \ .
\end{equation}
{\hfill \qed}
\begin{proposition}\label{Proposition-3}
If the form $\om^T=\omega^{(t_{1},\cdots ,t_{n})}$ 
satisfies the following equation,
\begin{equation}
\delta _{1}\cdots \delta _{n-1}\,\omega ^T + 
\delta_{n}\psi ^{T+E_1+\cdots+E_{n-1}-E_n}=0\ ,
\label{prop3}
\end{equation}
the general solution for it is given by
\begin{equation}
\omega ^{T}=\sum_{i=1}^{n}\delta _{i}\,\varphi^{T-E_i}\ .
\label{sol_3}
\end{equation}
\end{proposition}
{\bf Proof of Proposition \ref{Proposition-3}.}  Applying $\delta _{n}$ on
(\ref{prop3}) we have
\[
\delta _{1}\cdots \delta _{n-1}\delta _{n}\,\,\omega^T=0 \ ,
\]
whose solution is (\ref{sol_3}) by virtue of
Proposition \ref{Proposition-2}.
\qed
\begin{proposition}\label{Proposition-4}
If the form $\om^T=\omega^{(t_{1},\cdots ,t_{n})}$ 
satisfies the following set of equations
\begin{equation}
\delta _{i}\,\,\omega ^T+\delta _{n}\psi_{i}^{T+E_i-E_n}=0\ ,\quad
i=1,\cdots ,n-1\ ,  
\label{26}
\end{equation}
the general solution is given by
\begin{equation}
\omega ^T = \delta _{1}\cdots \delta _{n-1}\,
\varphi^{T-E_1-\cdots-E_{n-1}} + \delta_{n}\eta^{T-E_n}\ . 
\label{27}
\end{equation}
\end{proposition}
\setcounter{corollary}{0}
\begin{corollary}
Let $\om^T=\omega ^{(t_{1},\cdots ,t_{n})},$
obeys the following set of equations
\begin{equation}
\delta _{i}\omega^T=0\ ,\quad i=1,\cdots ,n\ .
\label{35}
\end{equation}
The general solution for (\ref{35}) is
\begin{equation}
\omega ^T=\delta _{1}\cdots \delta _{n}\,
\varphi^{T-E_1-\cdots-E_n} \ .  
\label{36}
\end{equation}
\end{corollary}
{\bf Proof of Proposition \ref{Proposition-4}.} 
From Proposition \ref{Proposition-1} we write the solution for the first
equation of the set (\ref{26}):
\begin{equation}
\omega ^T=\delta _{1}\varphi^{T-E_1}+\delta _{n}(\cdots)\ .
\label{28}
\end{equation}
Substituting it into the second equation of (\ref{26}) we have
\begin{equation}
\delta _{2}\delta _{1}\varphi^{T-E_1}+\delta_{n}(\cdots)=0\ .  
\label{29}
\end{equation}
From  Proposition \ref{Proposition-3} we get the solution for (\ref{29}):
\begin{equation}
\varphi ^{T-E_1}=\delta _{1}\varphi^{T-2E_1}
+\delta _{2}\varphi^{T-E_1-E_2} +\delta _{n}(\cdots)\ .
\label{30}
\end{equation}
Substituting (\ref{30}) into (\ref{28}) we arrive at
\begin{equation}
\omega^T = \delta_{1}\delta_{2}\varphi^{T-E_1-E_2}
+\delta _{n}(\cdots)\ .  
\label{31}
\end{equation}
Repeating the argument we finally obtain the result \equ{27}.
\qed

\section{Truncated extended forms  and cohomology}\label{app-c}

Let us consider as in Appendix \ref{app-b} forms 
$\f^R=\f^{(r_1,\cdots,r_n)}$, which may be fields or superfields. 
The notations and conventions are explained in the beginning of that 
appendix. The nilpotent extended operator $\tilde\d$ and
an extended $q$-form of total weight $q$ are defined as
\begin{equation}
\tilde\d =\dsum{i=1}{n}\d_i\ , \quad\quad 
\exfi^q = \dsum{R}{|R|=q}  \phi^R\ .
 \label{ext-form}
\end{equation}
Let us define the truncated extended $q$ forms associated to 
the heighest $H=(h_1,\cdots,h_n)$ 
(in short: truncated forms) as
\begin{equation}
 \trfi^q = \lc\exfi^q\rc^\truncation =
\dsum{R}{|R|=q,\,R \leq H}  \phi^R
 \label{truncatedform1}
\end{equation}
The truncation, indicated by the exponent ``$\truncation$'', means
discarding in the expression all the forms of 
degree not constrained by $R\le H$.
The polynomials in these truncated forms and their truncated
exterior derivatives span the space $\treh$, with the exterior
multiplication and derivation rules
\eq
\trfi_1^{q_1} \trfi_2^{q_2} = 
\lc \exfi_1^{q_1} \exfi_2^{q_2} \rc^\truncation\ ,\quad
\trde\trfi^q = \lc \tilde\d \exfi^q \rc^\truncation\ .
\eqn{trunc-alg}
$\trde$ is obviously nilpotent, and the operations defined in \equ{trunc-alg}
map $\treh$ to  $\treh$.
\begin{proposition}\label{Proposition-C}
The cohomology of $\trde$ in the space $\treh$
consists of the highest weight truncated forms 
\eq
\trfi^{|H|} = \f^H\ ,\quad 
\trfi^{|H|} \not=\trde\trfi^{|H|-1}\ .
\eqn{cohom-trde'}
\end{proposition}
{\bf Remark.} A highest weight truncated form $\trfi^{|H|}$ is always
closed: $\trde\trfi^{|H|}=0$.

\noindent{\bf Proof of Proposition \ref{Proposition-C}.}
The proof is by induction. The result being obvious for $n=1$, we shall
prove it for the generic case $n$ assuming it to hold for $n-1$. 
Let us divide the weights $r_1,\cdots,r_n$ in two subsets: $r_1$ and
$R'=(r_2,\cdots,r_n)$. We define accordingly the partially extended
operator $\exdpr$ and the partially extended forms
\eq
\exdpr = \dsum{i=2}{n}\d_i\ ,\qquad
\exfipr^{(q-r,r)} = \dsum{R'}{|R'|=r}  \phi^{(q-r,R')}\ ,\quad
r = 0,\cdots,q\ ,
 \label{par-ext-form}
\end{equation}
as well as the partially truncated extended forms
\eq
\trfipr^{(q-r,r)} =\lc \exfipr^{(q-r,r)} \rc^{\trunpr}
= \dsum{R'}{|R'|=r,\,R'\le H'}  \phi^{(q-r,R')}\ ,\quad
r = 0,\cdots,q\ ,
 \label{par-trunc-form}
\end{equation}
on which act the partially truncated derivative $\dpr$ defined by
\eq
\dpr\trfipr^{(q-r,r)} =\lc \exdpr\exfipr^{(q-r,r)} \rc^{\trunpr}
= \dsum{i=2}{n}\lp
 \dsum{R'}{|R'|=r,\,R'\le H'-E_i} \d_i \phi^{(q-r,R')} \rp \ ,\quad
r = 0,\cdots,q\ .
 \label{par-trunc-der}
\end{equation}
From the induction hipothesis, the cohomology of $\dpr$ 
is trivial in the subspace of the partially truncated forms
\equ{par-trunc-der} restricted by the condition $r<h'$~\footnote{We use the
notation $h'$ for $|H'|=\sum_{i=2}^n h_i$.}.

Let us now solve the cohomology equation
\eq
\trde\trfi^q= 0\ .
\eqn{trunc-coh-eq}
The truncated form \equ{truncatedform1}, can be written as
\begin{equation}
 \trfi^q = \dsum{r = {\rm Max}(0,q-h_1)}{{\rm Min}(q,h')}  
\trfipr^{{q-r,r}}  \ .
\label{truncatedform2}
\end{equation}
We have thus to examine separately the four cases $0\le q<h_1$, 
$h_1\le q<h'$, $h'\le|H|<h_1$, $q=|H|=h_1+h'$, corresponding 
respectively to the areas (1), (2), (3) and to the point (4) of
the figure \ref{diag-h1-H'}.
\setcounter{figure}{0}
\begin{figure}[ht]
\SetScale{0.8}\setlength{\unitlength}{0.8pt}  
\begin{center}\begin{picture}(300,250)(0,0)
\thicklines
\Vertex(0,0){1.5}
\Vertex(150,0){1.5}
\Vertex(0,200){1.5}
\Vertex(150,200){1.5}
\Line(0,0)(0,250)
\Line(0,0)(250,0)
\DashLine(0,200)(150,50){3}
\DashLine(0,150)(150,0){3}
\Line(0,200)(150,200)
\Line(150,0)(150,200)
\put(250,0){\vector(1,0){1}}
\put(0,250){\vector(0,1){1}}
\put(250,-12){$r_1$}
\put(150,-12){$h_1$}
\put(-18,200){$h'$}
\put(-15,250){$r'$}
\put(50,50){(1)}
\put(65,95){(2)}
\put(100,150){(3)}
\put(152,200){(4)}
\end{picture}\end{center}
\caption{Weight diagram for the partially truncated superforms
$\trfipr^{(r_1,r')}$. The numbers (1), (2), (3) and (4) refer to the 4 cases
examined in the text.}
\label{diag-h1-H'}
\end{figure}
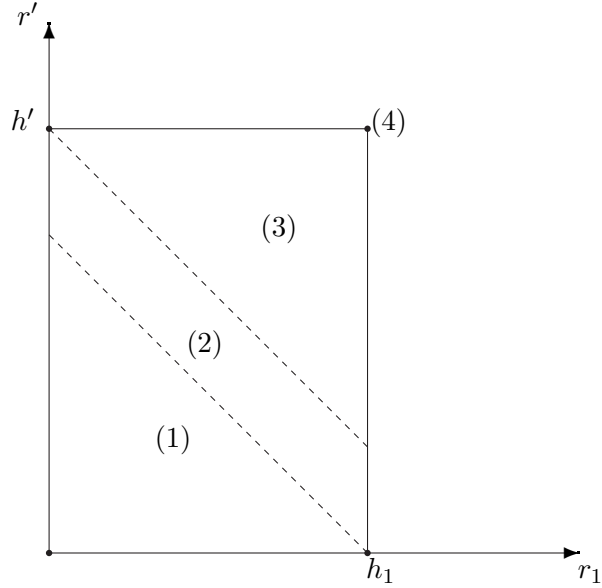


\noindent{\it Case 1}: $0 \le q < h_1$.\quad
We have
\[
\trfi^q = \dsum{r=0}{q}\trfipr^{(q-r,r)}  \ ,\qquad
\trde\trfi^q =  \dsum{r=0}{q} (\d_1+\dpr) \trfipr^{(q-r,r)} \ .
\]
The cohomology condition \equ{trunc-coh-eq} 
implies the following equations
\[\ba{l}
\d_1 \trfipr^{(q,0)} = 0 \ ,\es
\dpr \trfipr^{(q-r,r)} + \d_1 \trfipr^{(q-r-1,r+1)}  =  0 \ ,\quad
r= 1,\cdots,q-1        \ ,\es
 \dpr \trfipr^{(0,q)} = 0 \ .
\ea\] 
Solving these equations in turn, beginning from the first one, 
we obtain easily, using the triviality of the cohomology of $\d_1$:
\[\ba{l}
\trfipr^{(q,0)}  =  \d_1 \trpsipr^{(q-1,0)}\ , \es
\trfipr^{(q-r,r)}  =  \d_1 \trpsipr^{(q-r-1,r)} + \dpr \trpsipr^{(q-r,r-1)}\ , 
\quad r = 1, \cdots, q-1\ ,  \es
\trfipr^{(0,q)}  =  \dpr \trpsipr^{(0,q-1)}\ . 
\ea\] 
This result can be rewritten as
\eq
\trfi^q = \trde \trpsi^{q-1}\ ,\quad\mbox{with}\quad
\trpsi^{q-1} =  \dsum{r=0}{q-1} \trpsipr^{(q-1-r,r)} \ .
\eqn{case1}
We note that for $q=0$ we have $\trfi^0 = \trfipr^{(0,0)}$ and 
the solution is  $\trfi^0 = 0$.
We have thus proven the triviality of
the cohomology in the case 1.

\noindent {\it Case 2}: $h_1 \leq q < h' $.\quad
In this case the truncated form is given by
\[
\trfi^q = \dsum{r=q-h_1}{q}\trfipr^{(q-r,r)}  \ ,
\]
and the cohomology condition \equ{trunc-coh-eq} yields
\[\ba{l}
\dpr \trfipr^{(h_1-r, q - h_1+r)} 
+ \d_1 \trfipr^{(h_1-r-1,q-h_1+r+1)}  =  0\ ,\quad
r = 0,\cdots,h_1\ , \es
 \dpr \trfipr^{(0,q)}  =  0  \ ,
\ea\]
This time one has to begin with the last of these equations, and use the
induction hypothesis according to which
the  cohomology of $\dpr$ is trivial when applied to partial
truncated forms \equ{par-trunc-form} 
which are not of maximal weights, i.e such that $r<h'$. 
The solution reads
\[\ba{l}
\trfipr^{(0,q)}  =  \dpr \trpsipr^{(0,q-1)}  \ ,\es
\trfipr^{(r,q-r)} =
 \dpr \trpsipr^{(r,q-r-1)} + \d_1 \trpsipr^{(r-1,q-r)}\ ,
\quad  r = 1,\cdots,h_1\ , 
\ea\]
which again may be written as in \equ{case1}, showing the triviality of
the cohomology in the case 2. 

\noindent{\it Case 3}: $ h' \leq q < | H | $.\quad
The truncated form reads
\[
\trfi^q = \dsum{r=q-h_1}{q-h'}\trfipr^{(q-r,r)}  \ ,
\]
and the cohomology condition \equ{trunc-coh-eq} yields
\eq\ba{l}
\dpr \trfipr^{(h_1-r, q - h_1+r)} 
+ \d_1 \trfipr^{(h_1-r-1,q-h_1+r+1)}  =  0\ ,\quad
r = 0,\cdots,h_1+h'-q-1\ . 
\ea\eqn{c31}
The situation is a bit more subtle. We begin from the 
equation (\ref{c31}) with $r=0$, and 
use the result of Corollary 2 of
Proposition \ref{Proposition-1} -- valid due to the triviality of
the cohomology of both $\d_1$ and $\dpr$ --
from which we can write
\eq\ba{l}
\trfipr^{(h_1,q - h_1)} =  
  \dpr \trpsipr^{(h_1, q- h_1 -1)} + \d_1\trpsipr^{(h_1-1,q-h_1)} \ ,\es
\trfipr^{(h_1-1,q - h_1 +1)} =
  \dpr \trpsipr^{(h_1 -1,q - h_1)} + \d_1 \trpsipr^{(h_1-2,q-h_1+1)}\ .  
\ea\eqn{s32}
Now substituting this result in (\ref{c31}) for $r=1$ and using the
triviality of the cohomology of $\d_1$, we get the first of the
following equations -- the one for $r=2$:
\eq
\trfipr^{(h_1-r,q-h_1+r)} =
  \dpr \trpsipr^{(h_1-r,q-h_1+r-1)} + \d_1\trpsipr^{(h_1-r-1,q-h_1+r)}\ ,
\quad r = 2,\cdots,  h_1+h'-1\ .
\eqn{s33} 
The remaining ones, for $r\ge3$, are obtained in the usual way using the
triviality of the cohomology of $\d_1$.
The result \equ{s33} can be rewritten as \equ{case1}, 
showing the triviality of
the cohomology in the case 3.  

\noindent{\it Case 4}: $  q = | H | = h_1 + h' $\quad
This is the case of highest weight:
\[
\trfi^q = \trfipr^{(h_1, h')} = \phi^{(h^1, \cdots, h^n)}\ ,
\]
which satisfies identically the cohomology condition  
\equ{trunc-coh-eq}. It may be the $\dpr$-variation of some truncated
form $\trpsipr^{|H|-1}$, or not. In the latter case it belongs to the
cohomology of $\dpr$. 

Joining together the results of these four cases ends the proof 
of Proposition \ref{Proposition-C}.\qed

\section{Wess-Zumino gauge  and covariant superfields}\label{app-d}

As shown in~\cite{nos} it is possible to fix algebraically
the gauge degreees of 
freedom corresponding to the ghosts 
$c_{I_{1}...I_{n}}(x)$ ($1\leq n\leq \NT$), through the 
conditions\footnote{The field components of the superfields of the
theory are defined by (\ref{expansion-A}-\ref{expansions-E-C}).}
\begin{equation}
e_{I}(x)=0\ ,\quad e_{[II_{1}...I_{n}]}(x)=0\ (1\leq n\leq \NT)\ .  
\label{condi-NWZ}\end{equation}
This defines the so-called Wess-Zumino (WZ) gauge, analogous to the one
encountered in the supersymmetric Yang-Mills 
theories~\cite{1001}. 
We are left with the usual gauge degree of freedom corresponding to the ghost 
$c(x)$. The physical degrees of freedom are labeled by the covariant fields
\eq
\lac F_a,\,a_{I_1\cdots I_n},\,e^{(\rm M)}_{I_1\cdots I_{n+1}},\,
D_a a_{I_1\cdots I_n},\,D_a e^{(\rm M)}_{I_1\cdots I_{n+1}}
   \,;\,n\ge1\rac \ ,
\eqn{cov-WZ-basis}
where $F_a$ and $D_a$ are the Yang-Mills curvature and the covariant 
exterior derivative with respect to the connection $a$, and 
$e^{(\rm M)}_{I_1\cdots I_{n}}$ is the mixed symmetry tensor corresponding
to the second Young tableau in the right-hand side of the figure
\ref{Young}, and defined by the expansion
\setcounter{figure}{0}
\begin{figure}[ht]
\begin{center}\begin{picture}(150,80)(0,0)
\thicklines
\put(-30,12){$n+1$}
\put(-12,52){$2$}
\put(-12,42){$3$}
\Line(0,10)(10,10)
\Line(0,20)(10,20)
\Line(0,40)(10,40)
\Line(0,50)(10,50)
\Line(0,60)(10,60)
\DashLine(0,20)(0,40){2}
\DashLine(10,20)(10,40){2}
\Line(0,10)(0,20)
\Line(0,40)(0,60)
\Line(10,10)(10,20)
\Line(10,40)(10,60)
\put(18,32){$\times$}
\put(32,32){1}
\Line(40,30)(40,40)
\Line(50,30)(50,40)
\Line(40,30)(50,30)
\Line(40,40)(50,40)
\put(60,32){=}
\Line(80,0)(90,0)
\Line(80,10)(90,10)
\Line(80,20)(90,20)
\Line(80,40)(90,40)
\Line(80,50)(90,50)
\Line(80,60)(90,60)
\DashLine(80,20)(80,40){2}
\DashLine(90,20)(90,40){2}
\Line(80,0)(80,20)
\Line(80,40)(80,60)
\Line(90,0)(90,20)
\Line(90,40)(90,60)
\put(107,32){+}
\Line(130,10)(140,10)
\Line(130,20)(140,20)
\Line(130,40)(140,40)
\Line(130,50)(150,50)
\Line(130,60)(150,60)
\DashLine(130,20)(130,40){2}
\DashLine(140,20)(140,40){2}
\Line(130,10)(130,20)
\Line(130,40)(130,60)
\Line(140,10)(140,20)
\Line(140,40)(140,60)
\Line(150,50)(150,60)
\end{picture}\end{center}
\caption{Expansion \equ{mixed-symmetry} 
of the tensor $e_{I_1I_2\cdots I_n}$}
\label{Young}
\end{figure}
\eq\ba{ll}
&e_{I_1I_2\cdots I_{n+1}} = e_{[I_1I_2\cdots I_{n+1}]} 
  + e^{(\rm M)}_{I_1I_2\cdots I_{n+1}}\ ,\es
\mbox{with}\quad
&e^{(\rm M)}_{I_1I_2\cdots I_{n+1}} = \dfrac{1}{n+1}\dsum{k=2}{n+1}
\lp e_{I_1I_2\cdots I_k\cdots I_{n+1}}
  + e_{I_kI_2\cdots I_1\cdots I_{n+1}}\rp\ .
\ea\eqn{mixed-symmetry}
 The BRST transformations of the fields \equ{cov-WZ-basis}
are covariant:
\eq
\cas \f = -[c,\f]\ .
\eqn{BRS-WZ}

The stability of the WZ-gauge choice requires a redefinition 
of the SUSY operators -- acting on the fields \equ{cov-WZ-basis} and
$a$:
\eq
\tQ_I = Q_I + \d_{(\lambda_I)}\ ,
\eqn{def-q-tilde} 
where $\d_{(\lambda_I)}$ is a supergauge transformation of 
field dependent parameters
\eq
\la_{II_1\cdots I_n}= \frac{1}{n!n}e^{(\rm M)}_{II_1\cdots I_{n}}\ ,
\eqn{s-gauge-param}
equivalent to a superfield BRST transformation \equ{BRST-comp} with the
ghost components $c_{I_1\cdots I_{n}}$ given by 
\equ{s-gauge-param} and with $c=0$~\cite{nos}.
The superalgebra \equ{s-agebra} now closes on gauge
transformations
\eq
[\tQ_I,\tQ_J] = -2\d_{\rm gauge}(e^{(\rm M)}_{IJ}) \ .
\eqn{s-agebra-WZ}
of field dependent  parameters $e^{(\rm M)}_{IJ}$.

Let us now show that there is a bijection between the set of fields
\equ{cov-WZ-basis} and the following set of covariant superfields:
\eq
\lac F_A,\,\Psi^{(\rm A)}_{I_1\cdots I_n},\,
\F^{(\rm M)}_{I_1\cdots I_{n+1}},\,
D_A \Psi^{(\rm A)}_{I_1\cdots I_n},\,D_A \F^{(\rm M)}_{I_1\cdots I_{n+1}}
   \,;\,n\ge1\rac \ .
\eqn{curvature-basis}
$F_A$ and $D_A$ are the Yang-Mills curvature and the covariant 
exterior derivative with respect to the superfield connection $A$, and
the remaining elements are defined from the supercurvature components 
\equ{superComp} by
\eq\ba{l}
\Psi^{(\rm A)}_{I_1\cdots I_n}
 = \Psi_{\lc I_1|I_2\cdots I_n\rc}  \ ,\es
\F^{(\rm M)}_{I_1\cdots I_{n+1}} =
 \dfrac{n}{n+1} \lp \F_{I_1[I_2|I_3\cdots I_{n+1}]} 
                  + \F_{I_2[I_1|I_3\cdots I_{n+1}]} \rp \ .
\ea\eqn{def-cov-sfields}
The bracket $[\cdots ]$ means complete antisymmetrization in the indices,
the bar $|$ symbolizes covariant $\theta$-derivations:
\[
\Psi_{I_1|I_2\cdots I_n}= D_{I_2}\cdots D_{I_n}\Psi_{I_1}\ ,\quad 
D_{I} X = \pa_I X + [E_I,X]\ ,
\]
and the mixed symmetry tensors $\F^{(\rm M)}_{I_1\cdots I_{n+1}}$ 
belong to the expansion of the covariant $\theta$-derivatives 
$\F_{I_1I_2|I_3\cdots I_{n+1}}$ in irreducible
representations of the permutation group $S_{n+1}$: they correspond
to the Young diagram shown in the right-hand side of
the figure \ref{Young2}.
\begin{figure}[ht]
\begin{center}\begin{picture}(150,80)(0,10)
\thicklines
\put(-30,12){$n+1$}
\put(-12,52){$3$}
\put(-12,42){$4$}
\put(42,20){$1$}
\put(52,20){$2$}
\Line(0,10)(10,10)
\Line(0,20)(10,20)
\Line(0,40)(10,40)
\Line(0,50)(10,50)
\Line(0,60)(10,60)
\DashLine(0,20)(0,40){2}
\DashLine(10,20)(10,40){2}
\Line(0,10)(0,20)
\Line(0,40)(0,60)
\Line(10,10)(10,20)
\Line(10,40)(10,60)
\put(22,32){$\times$}
\Line(40,30)(60,30)
\Line(40,40)(60,40)
\Line(60,30)(60,40)
\Line(50,30)(50,40)
\Line(40,30)(40,40)
\put(75,32){=}
\Line(110,10)(120,10)
\Line(110,20)(120,20)
\Line(110,40)(120,40)
\Line(110,50)(120,50)
\Line(110,60)(130,60)
\DashLine(110,20)(110,40){2}
\DashLine(120,20)(120,40){2}
\Line(110,10)(110,20)
\Line(110,40)(110,70)%
\Line(120,10)(120,20)
\Line(120,40)(120,70)
\Line(130,60)(130,70)
\Line(110,70)(130,70)
\put(140,32){+ $\ \ \cdots$}
\end{picture}\end{center}
\caption{Expansion 
of the tensor $\F_{I_1I_2|I_3\cdots I_{n+1}}$}
\label{Young2}
\end{figure}

All the objects $X$ in \equ{curvature-basis} are covariant, i.e.
\[
\cas X = -[C,X]\ .
\]

The bijection between the set \equ{cov-WZ-basis} and 
the set \equ{curvature-basis}
is simply given by the fact that the elements of the former are equal to
the $\theta=0$ components of the elements of the latter, 
provided the WZ-gauge conditions \equ{condi-NWZ} are applied:
\eq\ba{lllll}
\Lp F_a,\,&a_{I_1\cdots I_n},\,&e^{(\rm M)}_{I_1\cdots I_{n+1}},\,&
D_a a_{I_1\cdots I_n},\,&D_a e^{(\rm M)}_{I_1\cdots I_{n+1}}\Rp 
=  \es
\Lp F_A,\,&\Psi^{(\rm A)}_{I_1\cdots I_n},\,&
\F^{(\rm M)}_{I_1\cdots I_{n+1}},\,&
D_A \Psi^{(\rm A)}_{I_1\cdots I_n},\,&
D_A \F^{(\rm M)}_{I_1\cdots I_{n+1}}\Rp 
\displaystyle{\biggr|}_{\te=0,\,{\rm WZ}}\ ,\quad n\ge1\ .
\ea\eqn{bijection}
This is obvious for $F_A$. For $\Psi^{(\rm A)}_{I_1\cdots I_n}$, we 
observe that each  $\theta$-derivative of
$\Psi_I$ brings in factors of $\theta$-derivatives  
$E_{I,I_1,\cdots I_k}$ or space-time covariant derivatives of
them. However, only completely antisymmetrized 
derivatives $E_{[I,I_1,\cdots I_k]}$ may contribute 
as factors to the completely
antisymmetric tensor $\Psi^{(\rm A)}$. Since these completely antisymmetrized 
derivatives vanish at $\theta=0$ due to the  WZ-gauge conditions, we are
left with the simple $\theta$-derivatives of $A$, which at $\theta=0$
yield the fields $a_{I_1\cdots I_n}$. The same conclusion holds  
for the terms $D_A \Psi^{(\rm A)}_{I_1\cdots I_n}$.  

The argument is similar for the terms $\F^{(\rm M)}_{I_1\cdots I_{n+1}}$ 
(and $D_A\F^{(\rm M)}_{I_1\cdots I_{n+1}}$): only factors of completely
antisymmetrized derivatives of $E_I$ may contribute to these mixed
symmetry tensors, made from covariant $\theta$-derivatives of the 
{\it symmetric} tensor $\F_{I_1I_2}$ and 
symbolized by the diagram shown in the right-hand side
of the figure \ref{Young2}. And this same diagram defines the 
symmetry properties of $e^{(\rm M)}_{I_1\cdots I_{n+1}}$.

\newpage


\end{document}